\documentclass[twocolumn,trackchanges]{aastex63}
\let\pwiflocal=\iffalse \let\pwifjournal=\iffalse
\usepackage[utf8]{inputenc}
\usepackage{amsmath,amssymb,gensymb,times,graphicx,morefloats,color}
\usepackage{wasysym}
\usepackage{textcomp}
\usepackage{afterpage, bm}
\usepackage{natbib,hyperref}
\usepackage{wrapfig}
\usepackage{float}
\usepackage{ulem}
\usepackage{xcolor}
\usepackage[outdir=./]{epstopdf}
\usepackage{mathrsfs}
\usepackage{mathtools}
\usepackage{url}

\usepackage{listings}
\definecolor{dkgreen}{rgb}{0,0.6,0}
\definecolor{gray}{rgb}{0.5,0.5,0.5}
\definecolor{mauve}{rgb}{0.58,0,0.82}
\definecolor{golden}{rgb}{0.86,0.65,0.01}
\newcommand{\teff}{T$_{\mathrm{eff}}$}
\newcommand{\logg}{log $g$}

\newcommand{\vmic}{v$_{\mathrm{micro}}$}

\begin{document}

\title{Distant Relatives: The Chemical Homogeneity of Comoving Pairs Identified in Gaia}
\author[0000-0003-3707-5746]{Tyler~Nelson}
\affiliation{Department of Astronomy, The University of Texas at Austin, 2515 Speedway Boulevard, Austin, TX 78712, USA}

\author[0000-0001-5082-9536]{Yuan-Sen Ting}
\affiliation{Institute for Advanced Study, Princeton, NJ 08540, USA}
\affiliation{Department of Astrophysical Sciences, Princeton University, Princeton, NJ 08544, USA}
\affiliation{Observatories of the Carnegie Institution of Washington, 813 Santa Barbara Street, Pasadena, CA 91101, USA}
\affiliation{Research School of Astronomy \& Astrophysics, Australian National University, Cotter Rd., Weston, ACT 2611, Australia}

\author[0000-0002-1423-2174]{Keith~Hawkins}
\affiliation{Department of Astronomy, The University of Texas at Austin, 2515 Speedway Boulevard, Austin, TX 78712, USA}

\author[0000-0002-4863-8842]{Alexander Ji}
\affiliation{Observatories of the Carnegie Institution of Washington, 813 Santa Barbara Street, Pasadena, CA 91101, USA}

\author[0000-0001-5625-5342]{Harshil Kamdar}
\affiliation{Harvard-Smithsonian Center for Astrophysics, Cambridge, MA, 02138, USA}

\author{Kareem El-Badry}
\affiliation{Department of Astronomy and Theoretical Astrophysics Center, University of California Berkeley, Berkeley, CA 94720}

\correspondingauthor{Tyler Nelson}
\email{tyler.nelson@utexas.edu}
\date{Accepted XX. Received YY; in original form ZZ}

\begin{abstract}

Comoving pairs, even at the separations of $\mathcal{O}(10^6)\,$AU, are a predicted reservoir of conatal stars. We present detailed chemical abundances of 62 stars in 31 comoving pairs with separations of $10^2 - 10^7\,$AU and 3D velocity differences $< 2 \mathrm{\ km \  s^{-1}}$. This sample includes both bound comoving pairs/wide binaries and unbound comoving pairs. Observations were taken using the MIKE spectrograph on the Magellan/Clay Telescope at high resolution ($\mathrm{R} \sim 45,000$) with a typical signal-to-noise ratio of 150 per pixel. With these spectra, we measure surface abundances for 24 elements, including Li, C, Na, Mg, Al, Si, Ca, Sc, Ti, V, Cr, Mn, Fe, Co, Ni, Cu, Zn, Sr, Y, Zr, Ba, La, Nd, Eu. Taking iron as the representative element, our sample of wide binaries is chemically homogeneous at the level of $0.05$ dex, which agrees with prior studies on wide binaries. Importantly, even systems at separations $2\times10^5-10^7\,$AU are homogeneous to $0.09$ dex, as opposed to the random pairs which have a dispersion of $0.23\,$dex. Assuming a mixture model of the wide binaries and random pairs, we find that $73 \pm 22\%$ of the comoving pairs at separations $2\times10^5-10^7\,$AU are conatal. Our results imply that a much larger parameter space of phase space may be used to find conatal stars, to study M-dwarfs, star cluster evolution, exoplanets, chemical tagging, and beyond.

\end{abstract}

\section{Introduction}
\label{introduction}

Stars which are born from the same gas, i.e., conatal, are critical in astronomy. The conatal nature of open clusters and wide binaries have made them indispensable laboratories for testing and improving our understanding of various areas of Galactic and stellar astrophysics. Both open clusters and wide binaries have been used to evaluate the current feasibility of chemical tagging \citep{DeSilva_07_open_clusters,Ting2012b,Ness2018,Andrews2019, wide_binaries_KH}. Open clusters are used to calibrate stellar parameter and chemical abundance pipelines for large surveys \citep[e.g.,][]{ASPCAP}. Wide binaries can be used to calibrate the metallicity of M-dwarfs \citep[e.g.,][]{Lepine_2007, Rojas_Ayala_2010, Montes_2018}, constrain the age-magnetic activity relation \citep[e.g.,][]{Garces_2011, Booth2017_age_magnetic_activity}, the age-metallicity relation \citep[e.g.,][]{Rebassa_2016}, and the initial-final mass relation \citep[e.g.,][]{Zhao_2012, Andrews_2015}. Besides, we can also use wide binaries to study exoplanet engulfment \citep[e.g.,][]{Melendez2017, Oh2018}.

Gravitationally bound binaries favor a conatal origin. For separations within a few thousand AU, binaries are mostly formed via turbulent core fragmentation \citep[][]{Offner_2010_star_formation, Lee2017_low_mass_binary}.  At separations up to $\sim 10^5\,$AU, wide binaries can be formed from dynamical unfolding of triple systems \citep{Reipurth_wb_3body}, the evaporation of star clusters \citep[][]{Kouwnhoven_wb_cluster_dissolving, Moeckel_wb_dissolving_cluster}, and chance gravitational capture of prestellar cores \citep{Tokovinin_wb_2017_grav_cores}. The predicted conatal nature of wide binaries motivated studies of their chemistry \citep{Andrews2019, Ramirez2019, wide_binaries_KH}. They have found wide binaries to be chemically homogeneous to $\sim 0.02-0.1$ dex in $\Delta\mathrm{[Fe/H]}$\footnote{Some systems have a larger metallicity difference \citep[e.g., $\Delta\mathrm{[Fe/H]}\sim0.2$ dex,][]{Oh2018}}. Nonetheless, the studies of chemical homogeneity for binaries remain rather limited ($\mathcal{O}(10)$ pairs) as large surveys often do not observe both stars due to the subsampling, and we have to resort to targeted studies. 

Comoving pairs, on the other hand, are even less studied than wide binaries. The investigation of detailed chemistry for comoving pairs remains largely non-existent for separations beyond a few parsecs. Comoving pairs separated by $\gtrsim 10^5\,$AU are no longer bound \citep{Jiang_wb_tidal_radius}. As outlined in \cite{Oh2017}, these unbound systems could arise from disrupted/dissolved conatal systems (open clusters, stellar binaries) or unrelated systems (chance alignments, gravitational resonance). Recently, simulations from \cite{Kamdar2018} argued comoving pairs have a high probability of being conatal provided that their 3D velocity difference ($\Delta\mathrm{v_{3D}}$) is below 2 km s$^{-1}$ at separations up to  $\mathcal{O}(10^6)\,$AU. Their simulations predicted that pairs with separations $\sim 10^6$ AU are $\sim80\%$ conatal. If true, this could dramatically expand the number of systems that could be conatal, which can improve stellar atmospheric models, study exoplanet engulfment, and calibrate surveys, among other applications. With the precise astrometry provided by the recent data release of Gaia \citep{Gaia_DR2, Gaia_eDR3} sufficient comoving systems can readily be located, and this study is set up to test this proposition -- {\em are comoving stars with large separations conatal?}

A key characteristic of conatal stars is their chemical composition’s homogeneity; conatal stars are expected to have similar initial chemical compositions as they are usually formed from well-mixed ISM gas \citep{Feng_2014}. Nonetheless, measured surface abundances can be changed by internal processes \citep[atomic diffusion, rotational mixing, dredge-up, gravitational settling, radiative levitation,][]{Rotational_mixing,atomic_diff}, model limitations \citep[incomplete laboratory data, NLTE, 3D][]{NLTE_Ruchti2013,benchmark_temperature_grav, Nissen_rev2018, Jofre_rev2019}, and methodological limitations \citep[][]{ benchmark_metal, Jofre_rev2019}. All these effects could complicate any interpretation of the chemical homogeneity of the two stars.

To distinguish model systematic from astrophysical processes, in this study, we will appeal to differential analysis of stellar twins -- i.e., stellar pairs with similar stellar parameters. Differential abundances can remove many of these effects provided the stars are close in stellar parameters because the systematic effects in the star of interest and reference star largely cancel out \citep[see][]{gray_book, Nissen_rev2018, Jofre_rev2019}.

In this study, we measure abundance differences in 24 elements for 33 comoving pairs with $\Delta\mathrm{v_{3D}} < 2 \ \mathrm{km \ s^{-1}}$ and separations between $\sim 10^2 - 10^7\,$AU. With access to homogeneous high resolution, high signal-to-noise data, we uniquely positioned to explore the detailed abundances from the wide binaries and the comoving stars and the chemical homogeneity of the two components. In Section \ref{target_selection} we describe the observations. Section \ref{bacchus_section} details the methods for estimating stellar parameters and abundances. Our results are presented in Section \ref{Results}, in which we explore the conatal fraction of the comoving pairs. These results are discussed in Section \ref{discussion} and summarized in Section \ref{summary}.

\section{Data Properties}
\label{target_selection}

Our target selection focused on comoving pairs of FG dwarfs within 300 pc of the Solar neighborhood. FG dwarfs were selected because they are both luminous, and their spectral models are best characterized. We create the initial list of sources with the following ADQL query:
\begin{lstlisting}[deletekeywords={DEC, DECIMAL, DECLARE}]
SELECT *
FROM gaiadr2.gaia_source
WHERE ra > 165
OR ra < 15
AND dec < 25 
AND bp_rp BETWEEN 0.5 AND 1.5
AND 1000/parallax < 300
AND parallax_over_error > 5
AND phot_g_mean_mag IS NOT NULL
AND ((phot_g_mean_mag - 5*LOG10(1000/parallax) + 5) BETWEEN 2 AND 7)
AND radial_velocity IS NOT NULL
\end{lstlisting}

Our range of $\mathrm{bp\_rp}$ was selected to exclude hot stars as some of them could be rapid rotators with broad spectral features which complicate spectral analysis. Next, to ensure that our systems are genuine (isolated) comoving stars, we filter out members of open clusters and stellar associations with a friend-of-friend (FoF) connectivity cut following \cite{kamdar_streams}, removing any pairs that belong to an aggregate with more than 4 FoF companions (with separations $< 10\,$ pc, $\Delta{\rm v_{3D}} < 5\,$ km s$^{-1}$). We verify the efficacy of this filtering by cross-matching our sample with the open clusters and comoving groups catalogs from \cite{Kounkel_2019} and \cite{Cantat_Gaudin_2020}, finding no overlap. From this list of possible sources, we calculate the 3D velocity separation $\Delta\mathrm{v_{3D}}$ and spatial separation using Galpy \citep{Galpy} using astrometry and radial velocities measurements from Gaia DR2. The target selection prioritized stars in similar locations on the color-magnitude diagram (CMD) to reduce the influence of systematic errors on the derived elemental abundances. Together, we consider the comoving pairs that satisfy the following criteria:
\begin{itemize}
    \item $|\Delta\mathrm{M}_{\rm G}| < 0.5\mathrm{\ mag}$
    \item $|\Delta(\mathrm{G_{BP} - G_{RP}})| < 0.1 \mathrm{\ mag}$ 
    \item $\Delta\mathrm{v_{3D}} < 2 \mathrm{\ km \ s^{-1}}$ 
    \item 3D separation $ < 300 \mathrm{\ pc}$
\end{itemize}

The remaining comoving pairs were binned in log separation and sorted by apparent magnitude. The final sample was selected by taking the most luminous $\sim10$ pairs from each log separation bin. 

Throughout this study, we will use the term comoving stars to signify both bound and unbound pairs of stellar companions. In particular, for consistency, we will use the words ``close" comoving pairs and wide binaries interchangeably to refer to (mostly bound) wide binaries and ``far'' comoving pairs to be the primarily unbound systems. For simplicity, we define close comoving pairs as those with 3D spatial separation of 1 pc (i.e.,  $2\times 10^5\,{\rm AU}$), and the others far comoving pairs. This choice is partly motivated by the fact that in simulations from \citet{kamdar2019b}, the minimum separation of unbound comoving pairs have a separation of 1 pc. This is also consistent with the previous study from \citet{Jiang_wb_tidal_radius}.

We observed 33 pairs of comoving stars using the MIKE spectrograph on the \emph{Magellan/Clay} telescope \citep{Bernstein03} from June 13 to June 16, 2019. These 33 pairs constitute our ``main sample''. We also observe two additional pairs with $\Delta\mathrm{v_{3D}} > 2\mathrm{\ km \ s^{-1}}$ as a control sample. But unless otherwise stated, we will only refer to the main sample throughout this study. Additionally, we also observed four Gaia benchmark stars \citep{benchmark_metal, benchmark_library, benchmark_temperature_grav} dispersed over the sampled CMD to improve the data reduction. The benchmark stars used were 18 Sco, bet Vir, HD 140283, and mu Ara. The instrument employs a blue and red spectrograph to cover 3350--5000 \AA \ and 4900--9500 \AA \ respectively. We used the 0.5'' slit with $2 \times 1$ binning, which gave the blue and red spectrographs typical resolving powers of 50,000 and 40,000, respectively. The median signal-to-noise ratio (SNR) per pixel for the blue and red chips were 121 and 185. The observational details are given in Table \ref{tab:obs_table}. 

\begin{figure}
    \centering
    \includegraphics[width=1\columnwidth]{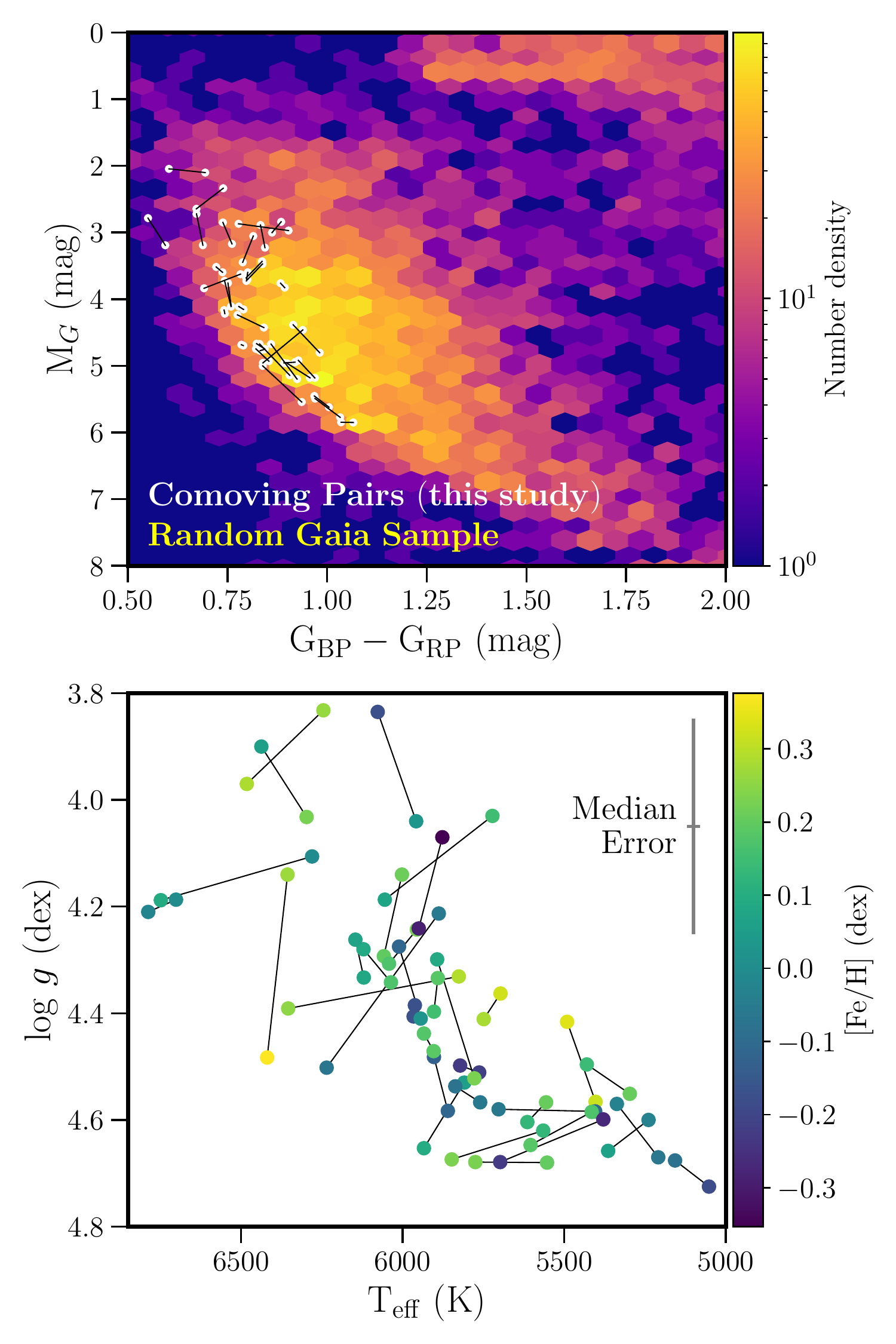}
    \caption{Top panel: The observed stars (white circles) are overplotted on 200,000 random stars from Gaia DR2 (background). Most of the samples are main-sequence stars because giant-giant stellar twins are rare, and we focus on the bright local sample. Bottom panel: Stellar parameter fits for the comoving pairs from BACCHUS. Representative (median) error bars are given in the upper right corner of the plot. For each pair, the two comoving components are connected by a black line.}
    \label{fig:gaia_CMD}
\end{figure}

Our sample size of 33 pairs is comparable to previous detailed chemical studies of wide binaries \citep[][]{Andrews2019, Ramirez2019, wide_binaries_KH}, but expands the range of separations by two orders of magnitude. Previous studies mostly restrict their sample to $\lesssim 10^5\,$AU, whereas in this study we have comoving pairs with separations up to $10^7\,$AU. In the top panel of Figure \ref{fig:gaia_CMD} we show the CMD of our comoving targets (white circles) with a random sample of 200,000 stars from Gaia DR2 plotted in the background for reference. 

Among our sample, we found that, for two particular pairs, each of which had a component that exhibit large deviations in the radial velocity measurements from Gaia eDR3\footnote{Note that the radial velocity in eDR3 follows the one from DR2} and our RV derived with the MIKE spectra. Both systems were flagged as outliers based on the interquartile range (IQR) test\footnote{Consider a data distribution, with $P_{i}$ denoting the ith percentile of said distribution. A data point $D$ is labeled an outlier if $D$ is outside the interval $P_{50} \pm 1.5\times (P_{75} - P_{25})$, where $(P_{75} - P_{25})$ is the interquartile range.}. This deviation could indicate an unresolved companion star. We exclude them from the primary sample because the unresolved binaries in hierarchical triplets can bias metallicity estimates by 0.1 dex \citep[][]{unresolved_binaries_apogee_effects}. Excluding these pairs leaves us with 31 pairs of comoving stars. 

\begin{table*}
\caption{Abbreviated target list for the 33 comoving pairs investigated in this study. A complete machine-readable version is available online. For compactness, we only display a subset of the columns. Comoving pairs can be determined using the identifier column. The fourth and fifth columns are the parallax and parallax errors from Gaia eDR3.}
    \label{tab:obs_table}
\centering

\begin{tabular}{cccccccccccc}
\hline
\hline
Gaia eDR3 ID & Identifier & RA & DEC & $\varpi$ & $\sigma_{\varpi}$ & G & BP - RP & RV & $\sigma_{\mathrm{RV}}$ & SNR$_{\mathrm{Blue}}$ & ... \\

& & ($^o$)& ($^o$)& $\mathrm{(mas)}$ & $\mathrm{(mas)}$& $\mathrm{(mag)}$ &$\mathrm{(mag)}$ & $\mathrm{(km \ s^{-1})}$ & $\mathrm{(km \ s^{-1})}$ &  & ...\\
\hline
6160638833832941312 & CM00A & 191.8605 & -30.8986 & 9.931 & 0.017 & 4.75 & 0.84 & -12.50 & 0.05 & 86 & ... \\
3471119180522314624 & CM00B & 189.8021 & -29.1721 & 9.613 & 0.014 & 4.77 & 0.83 & -11.23 & 0.06 & 75 & ... \\
3613454637229633664 & CM01A & 206.5261 & -11.7535 & 4.092 & 0.015 & 3.45 & 0.79 & 22.96 & 0.03 & 93 & ... \\
3613454637229634176 & CM01B & 206.5178 & -11.7507 & 4.085 & 0.016 & 3.05 & 0.81 & 22.17 & 0.03 & 120 & ... \\
3639520621950395904 & CM02A & 215.6611 & -7.7688 & 23.742 & 0.025 & 4.16 & 0.74 & -32.62 & 0.03 & 209 & ... \\
3639520621950395776 & CM02B & 215.6613 & -7.7704 & 23.802 & 0.024 & 4.22 & 0.74 & -33.26 & 0.03 & 167 & ... \\
5897704951769034368 & CM03A & 218.3675 & -52.5846 & 5.880 & 0.017 & 3.47 & 0.84 & 43.98 & 0.03 & 114 & ... \\
5897785667103983616 & CM03B & 217.2034 & -52.9914 & 5.645 & 0.018 & 3.71 & 0.80 & 44.24 & 0.03 & 94 & ... \\
6224633983987510528 & CM04A & 225.8985 & -27.8432 & 19.433 & 0.027 & 4.70 & 0.79 & -15.77 & 0.03 & 131 & ... \\
6224633983987511552 & CM04B & 225.9000 & -27.8416 & 19.413 & 0.029 & 4.68 & 0.78 & -14.99 & 0.03 & 145 & ... \\
\hline
\end{tabular}
\end{table*}

\begin{figure*}
    \centering
    \includegraphics[width=2\columnwidth]{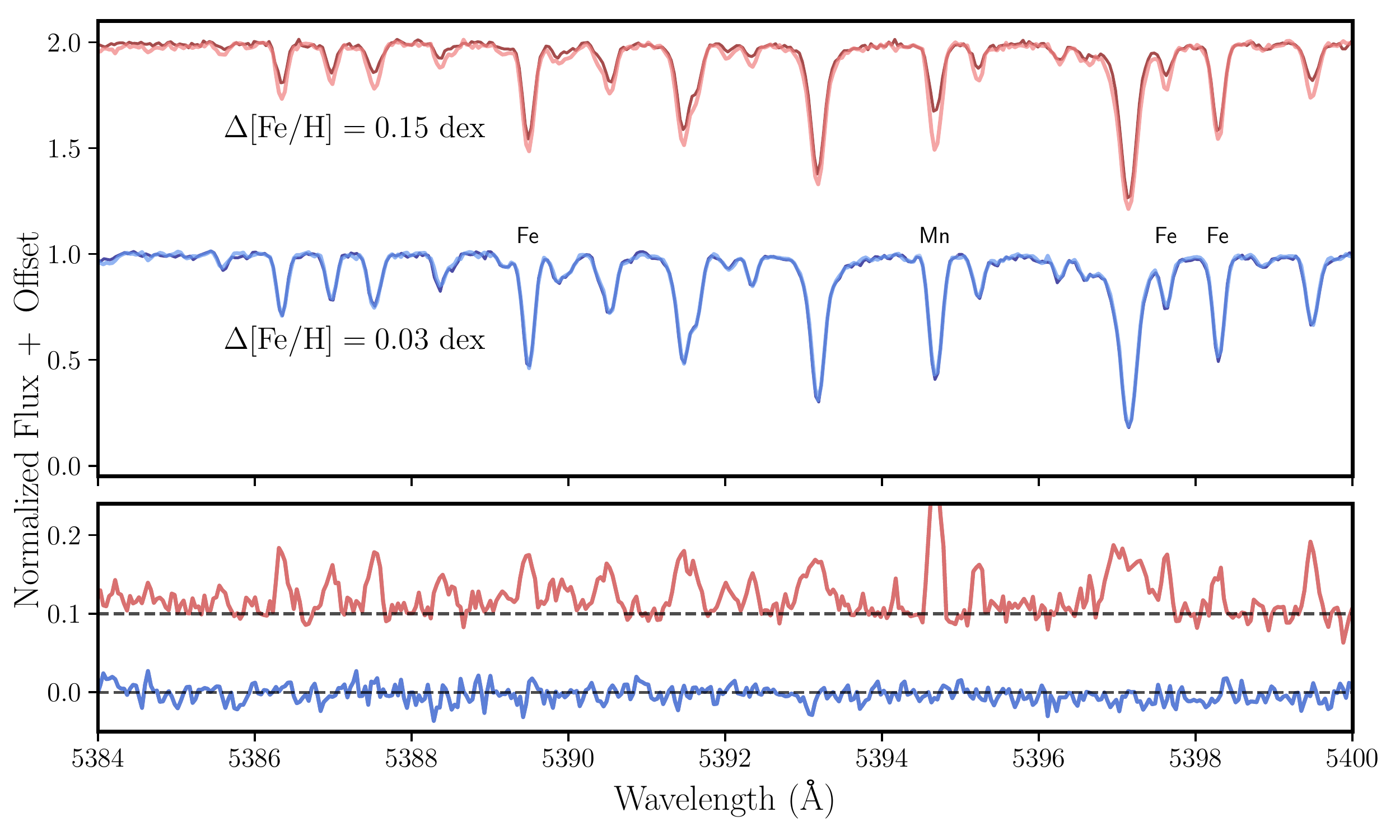}
    \caption{
    The upper panel shows continuum normalized spectra of both components of two comoving pairs offset by a constant with $\Delta$[Fe/H] of 0.15 dex and 0.03 dex for the upper and lower set of spectra, respectively. Spectral lines which passed the initial quality cut (see Section \ref{bacchus_section}) are also shown in the upper panel. The lower panel shows the difference between the components. Since we only target stellar twins with $\Delta\mathrm{T_{eff}} \lesssim 100 \  \mathrm{K}$ for both comoving pairs, the differences in the spectra are primarily due to the element abundances of the stars.}
    \label{fig:sample_spectrum}
\end{figure*}

\begin{figure}
    \centering
    \includegraphics[width=1\columnwidth]{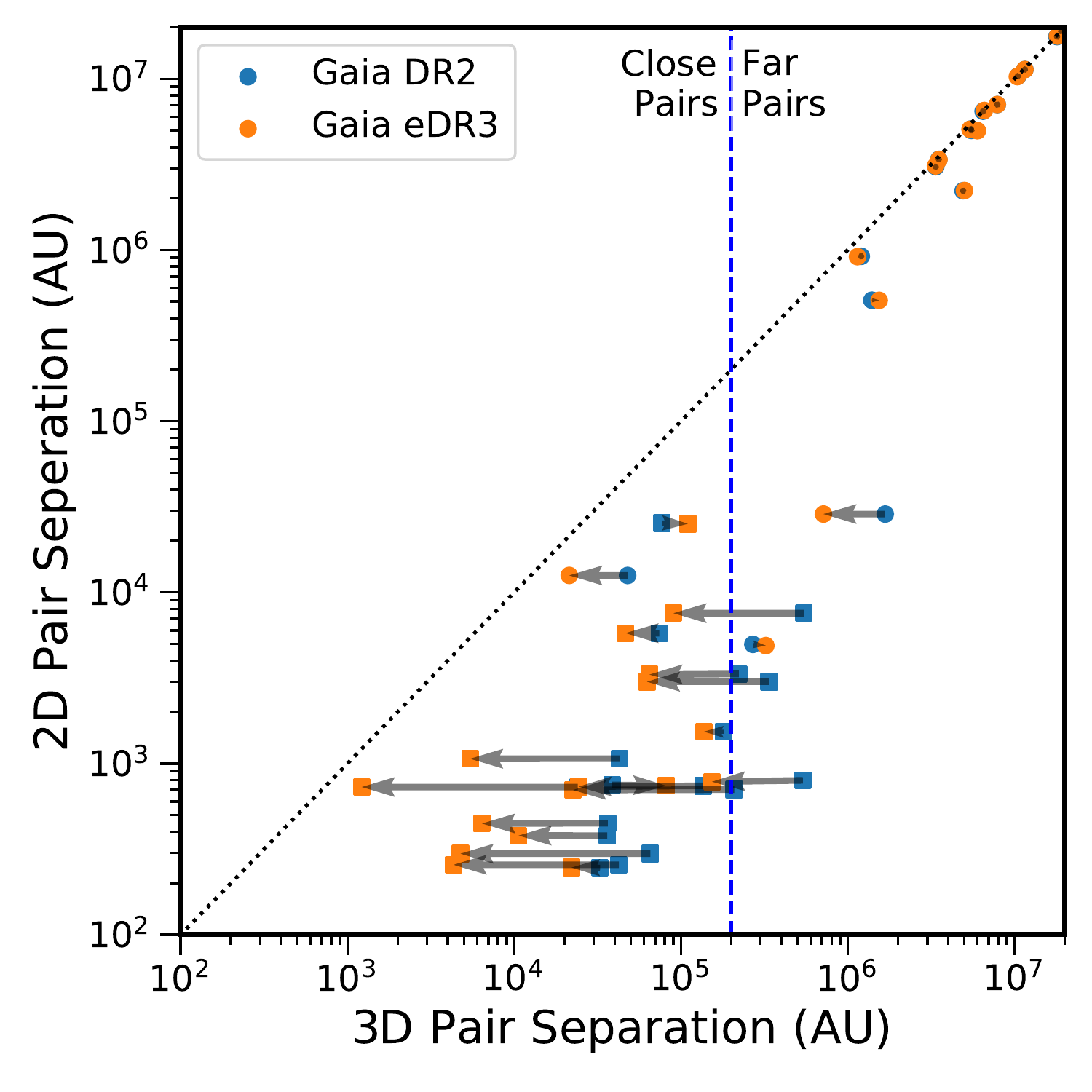}
    \caption{The 2D projected pair separation plotted against the 3D pair separation for both Gaia DR2 (blue) and eDR3 (orange). The square symbols indicate pairs which are in the wide binary catalog from \cite{elbadry_wb_edr3_preprint}, and circles are used otherwise. All separations are calculated directly from Gaia eDR3 astrometry. For a given pair, the grey arrow links the DR2 data to the eDR3 calculation. The blue dashed line indicates our defined boundary separating the close and far comoving pairs. For the far comoving pairs, the parallax error is subdominant. As such, the 2D separations and 3D separations line up. However, for close moving pairs, the parallax error dominates over the separation calculation, and the calculated 3D separation is much larger than the 2D projected separation. Since bound binaries are expected to have smaller separations ($\lesssim 10^5\,{\rm AU}$), this further validates that the parallax errors dominate for the close comoving pairs. Due to this limitation, we impose a geometric prior (Appendix~\ref{sec:3d_WB}) when calculating the 3D separation for the close comoving pairs.}
    \label{fig:separation_prior_motivation}
\end{figure}

We reduce the raw MIKE data with CarPy \citep{Kelson03} which outputs wavelength-calibrated multi-order echelle spectra for the science targets and flats.  After flat-fielding, the spectra were normalized by fitting a cubic spline as a pseudo continuum for each order. We determine the pseudo continuum by iterative sigma clipping. We discard 100 pixels on both ends of each order because the blaze function is less well defined due to the lack of signal. Orders are then combined using a flux error weighted average. The sigma clipping process was tuned by comparing observations of four Gaia Benchmark stars to the online library \citep{benchmark_library}. We employ iSpec \citep{ispec} to determine the radial velocity (RV) correction and remove cosmic rays. 

The reduced spectra for two comoving pairs in the wavelength regime of $5384 \le \lambda \le 5400 \mathrm{\ \AA}$ are shown in Figure \ref{fig:sample_spectrum}. In the top panel, we compare the spectra of both components for the two comoving pairs. The bottom panel shows the difference in spectra between the comoving components. Since we only target stellar twins, the difference predominantly comes from the difference in elemental abundances. The two spectra exhibit only minimal differences when they have similar chemistry. The careful selection of stellar twins allows us to derive elemental abundances with high fidelity for this homogeneous sample of comoving pairs.

\subsection{Positions and Velocities}

\label{separations_velocities_method1}
We calculate the positions and velocities for each star using Gaia eDR3 astrometry and spectroscopic RVs from our observations. Note that we made the target selection with DR2 before eDR3 became available. However, since Gaia eDR3 improves Gaia DR2 astrometry, we decide to use eDR3 astrometry for the final determination and plots throughout this study. This update causes some objects within our main sample to appear outside the $\Delta\mathrm{v_{3D}} < 2 \ \mathrm{km \ s^{-1}}$ cut, even though our our main sample are selected to be within this boundary with the DR2 values.

We will consider two approaches to calculate the 3D separations between members of comoving pairs. The first (``method 1'') uses the 6D phase space information to calculate the 3D separation directly using a Monte Carlo approach. We construct a multivariate normal distribution for each star with mean of the reported values and a covariance matrix using the 5D measured correlations in astrometry from Gaia eDR3 as well as the RV errors from the iSpec RV fit. We sample from this distribution 10,000 times and apply a transformation from equatorial to Galactic Cartesian coordinates to each sample. Finally, we calculate the 3D separation and velocity difference of the samples in Cartesian coordinates, taking the mean and standard deviation across samples as the point estimate and uncertainty for these quantities.

This method does not invoke additional assumptions about the geometry or separation distribution of pairs. However, for comoving pairs with 3D separations, $\lesssim 10^5\,{\rm AU}$, the uncertainty in 3D separation is often significantly inflated by parallax uncertainties, making it impossible to accurately measure 3D separations using method 1. Consider two stars with identical true parallax, $\varpi$, and parallax uncertainty, $\sigma_{\varpi}$. In the limit of $\sigma_{\varpi}/\varpi \ll 1$, the typical difference in {\it apparent} line-of-sight distance between the two stars is $\Delta d_{{\rm apparent}}\sim\frac{1000\,{\rm pc}}{\left(\varpi/{\rm mas}\right)}\times\sigma_{\varpi}/\varpi$. For typical pairs in our sample, with $\varpi \sim 10\,\rm mas$ and $\sigma_{\varpi}\sim 0.02$ mas, this translates to an apparent distance difference of $\sim 4\times 10^4$\,AU, which represents the amount by which the distance differences of typical pairs will be inflated by parallax uncertainties. For the pairs in our sample with the smallest $\varpi$ and largest $\sigma_{\varpi}$, this value is $\sim 3\times 10^5$\,AU. This intuition also prompted us to divide the close comoving pairs and the far comoving pairs with a 1pc (or $2 \times 10^5\,$AU) threshold.

It is expected that many of the closest pairs in the sample are gravitationally bound wide binaries with true 3D separations smaller than this distance ``resolution'' afforded by the {\it Gaia} parallaxes. Parallax uncertainties effectively stretch these pairs out along the line of sight, in an effect that is analogous to the ``fingers of God” distortion in cosmological surveys, where galaxy pairs and clusters are stretched out by redshift uncertainties. For such pairs, 
the point estimate of the 3D separation from method 1 will dramatically overestimate the true 3D separation. The {\it projected} separation (which is not subject to inflation by parallax uncertainties) will be much smaller than the point estimate of the 3D separation. 



Figure \ref{fig:separation_prior_motivation} compares the 2D and apparent 3D pair separations for all 33 comoving pairs. The 3D separations are calculated using method 1, as described above. We also show the results separately using astrometry from DR2 and eDR3. At large separations, where the parallax uncertainties are small compared to the parallax differences, the 2D and 3D separations agree within a factor of a few, and difference between the separations calculated using DR2 and eDR3 astrometry are minor. By contrast, pairs with 2D separations less than $10^5$\,AU fall far below the one-to-one line, with projected separations that are much smaller than the calculated 3D separations. For a majority of these pairs, the apparent 3D separation shrinks between DR2 and eDR3 astrometry. This reflects the fact that parallax uncertainties are smaller in eDR3 data, causing the separation inflation to be less severe. Cross-matching with \cite{elbadry_wb_edr3_preprint} (E21) shows that almost all of these pairs have a high probability of being gravitationally bound wide binaries,\footnote{In particular, 18 of the 21 pairs in our sample with projected 2D separations below $10^5$\,AU are in the E21 catalog; these all have a reported chance alignment probability of less than 0.001. None of the pairs with 2D separations above $10^5$\,AU are in the catalog. The three pairs that have 2D separations below $10^5$\,AU (and a 3D separation of $\sim 10^5-10^6\,$AU) and are not in the catalog only narrowly fail the cuts on parallax and/or proper motion consistency required for membership in the E21 catalog and thus may still be bound.} suggesting that their 3D separations are indeed over-estimated due to the parallax errors. 

This suggests that the separations of the close comoving pairs are dominated by parallax errors and that even Gaia eDR3 cannot ``resolve'' their line-of-sight distance difference. Therefore, for close pairs, we instead derive constraints on the 3D separation based on the projected separation, the assumption of random viewing angles, and a prior on the 3D separation distribution of wide binaries. This is describe in Appendix~\ref{sec:3d_WB}. We refer to this method of estimating the 3D separation, which only relies on the projected 2D separation, as ``method 2''. For method 2, the point estimate and 1$\sigma$ confidence interval on the 3D separation, $r_{\mathrm{3D}}$ are $r_{\mathrm{3D}}=1.12_{-0.11}^{+0.49}\times r_{\mathrm{2D}}$, where $r_{\mathrm{2D}}$ is the projected separation. Motivation for this expression is given in Appendix~\ref{sec:3d_WB}.

With different separation estimators for close pairs (thought to be binaries) and far pairs (thought to be unbound), it is necessary to decide where to draw the line between the two regimes. For simplicity, we use method 2 on all pairs for which the 3D separations calculated using method 1 are less than $2\times 10^5$\,AU (dashed blue line in Figure~\ref{fig:separation_prior_motivation}). It is possible that a few pairs wider than this (i.e. with $10^5-10^6\,$AU), which also have projected separations significantly smaller than their calculated 3D separations, are also binaries. Improved parallax uncertainties in future {\it Gaia} data releases will help determine whether they are bound. But we have checked that separating the bound and unbound pairs with a $10^6\,$AU threshold leaves most of our qualitative results unaltered. 

As most studies usually operate at the projected space instead of the 3D space adopted in this study, when comparing with the literature values (\cite{Ramirez2019}, and \cite{wide_binaries_KH}), we calculate the 3D separations and velocity differences of their targets with the same process as detailed above. We adopt the latest Gaia eDR3 astrometry and the RV from the literature. RV data was not provided in \cite{Ramirez2019}. Hence we used Gaia eDR3 RVs for the \cite{Ramirez2019} data set.

\begin{deluxetable*}{cccccccccc}
\tablecaption{The 3D separations ($\mathrm{S}$) and velocity differences ($\Delta\mathrm{v_{3D}}$) between comoving pairs. A full machine-readable version is available online. The subscript eDR3 indicates separations calculated solely from Gaia eDR3 astrometry and the spectroscopic RV. As discussed in Section \ref{separations_velocities_method1}, the separations calculated this way may overestimate the true separation between the close comoving components. For the subset of comoving pairs suspected to be wide binaries, we invoke a geometric prior based to infer the 3D separations based on the better measured 2D projected separations (Appendix~\ref{sec:3d_WB}). We only perform this correction for the close comoving pairs. These separations are denoted with a subscript ``geometric''. The uncertainty for the geometric separations is asymmetric, so the upper and lower error bars (16 and 84 percentiles) for this separation are listed separately.  The results in this paper assumes $\mathrm{S_{geometric}}$ for close comoving pairs (i.e., $\mathrm{S_{eDR3}} < 2\times10^5$ \ AU) and $\mathrm{S_{eDR3}}$ for the far comoving pairs. \label{tab:kinematic_table}}

\tablehead{Gaia eDR3 ID for &  Gaia eDR3 ID for & Identifier & $\mathrm{S_{eDR3}}$ & $\sigma_{\mathrm{S_{eDR3}}}$ & $\Delta\mathrm{v_{3D}}$ & $\sigma_{\Delta\mathrm{v_{3D}}}$ & $\mathrm{S_{geometric}}$ & $\sigma_{\mathrm{lower}_{\mathrm{S_{geometric}}}}$ & $\sigma_{\mathrm{upper}_{\mathrm{S_{geometric}}}}$ \\
Component A & Component B & & (AU) & (AU) & (km s$^{-1}$) & (km s$^{-1}$) & (AU) & (AU) & (AU)}
\startdata
6160638833832941312 & 3471119180522314624 & CM00 & 1.1 x 10$^{6}$ & 2.7 x 10$^{4}$ & 0.86 & 0.05 & -- & -- & -- \\
3613454637229633664 & 3613454637229634176 & CM01 & 9.0 x 10$^{4}$ & 1.8 x 10$^{5}$ & 1.02 & 0.08 & 8.5 x 10$^{3}$ & 8.3 x 10$^{2}$ & 3.7 x 10$^{3}$ \\
3639520621950395904 & 3639520621950395776 & CM02 & 2.2 x 10$^{4}$ & 9.3 x 10$^{3}$ & 1.29 & 0.04 & 2.8 x 10$^{2}$ & 27 & 1.2 x 10$^{2}$ \\
5897704951769034368 & 5897785667103983616 & CM03 & 1.5 x 10$^{6}$ & 1.0 x 10$^{5}$ & 0.93 & 0.05 & -- & -- & -- \\
6224633983987510528 & 6224633983987511552 & CM04 & 1.1 x 10$^{4}$ & 1.6 x 10$^{4}$ & 1.71 & 0.08 & 4.2 x 10$^{2}$ & 41 & 1.9 x 10$^{2}$ \\
6004256909931771136 & 6004256802548180736 & CM05 & 1.2 x 10$^{3}$ & 8.6 x 10$^{3}$ & 0.82 & 0.06 & 8.2 x 10$^{2}$ & 80 & 3.6 x 10$^{2}$ \\
5877155289930033024 & 5877155289930029184 & CM06 & 6.5 x 10$^{4}$ & 6.7 x 10$^{4}$ & 0.70 & 0.06 & 3.7 x 10$^{3}$ & 3.6 x 10$^{2}$ & 1.6 x 10$^{3}$ \\
5798991008295109120 & 5798991008295120896 & CM07 & 3.2 x 10$^{5}$ & 7.1 x 10$^{4}$ & 1.43 & 0.07 & -- & -- & -- \\
6214634887804201728 & 6214634887804233088 & CM08 & 6.4 x 10$^{3}$ & 4.8 x 10$^{4}$ & 0.58 & 0.04 & 5.0 x 10$^{2}$ & 49 & 2.2 x 10$^{2}$ \\
6208919660724640896 & 6208919660723526272 & CM09 & 8.1 x 10$^{4}$ & 3.2 x 10$^{4}$ & 0.79 & 0.06 & 8.3 x 10$^{2}$ & 81 & 3.6 x 10$^{2}$ \\
\enddata
\end{deluxetable*}

\section{Stellar Parameters and Abundances}
\label{bacchus_section}

We use the Brussels Automatic Code for Characterizing High accUracy Spectra \citep[BACCHUS,][]{bacchus2016} to determine stellar parameters (i.e., \teff, \logg, [Fe/H], \vmic) and abundances. An overview of the software is provided below. For a more comprehensive description we refer the reader to Section 2.2 of \cite{BACCHUS_kh_description}, \cite{bacchus2016}, and Appendix A.12 of \cite{GaiaESO_UVES_2014}.

BACCHUS fits observations using spectral synthesis. These spectra were constructed from MARCS model atmospheres \citep{MARCs} using \texttt{TURBOSPECTRUM} \citep{turbospectrum} for radiative transfer. MARCS models are calculated in 1D LTE. If the surface gravity (\logg) is $\ge 3.0$ dex, plane-parallel models are used, and spherical models otherwise. The atmospheric composition uses solar abundance from \cite{grevesse_solar_abunds} scaled by metallicity for most elements.  MARCs models use separate abundance estimates for C, N, and O \citep[see][Section 4 for more information]{MARCs}\footnote{Some MARCs models take into account the Galactic trends in [$\alpha$/Fe] vs [Fe/H]. These models are used for $\mathrm{[Fe/H]} \le -1$ dex. However, over the metallicity range of $0\pm0.3$ dex spanned by this sample, such effect is minimal. We thus assume the solar scaling for the $\alpha$-captured elements for this study.}. We use Gaia-ESO line list version 5 \citep{Heiter2019} for atomic transitions. The line list includes hyperfine structure splitting for Sc I, V I, Mn I, Co I, Cu I, Ba II, Eu II, La II, Pr II, Nd II, Sm II. We also include molecular data for CH \citep{CH_data_14}, C$_2$, CN, OH, MgH (T. Masseron, private communication), SiH \citep{Kurucz_atomic_molecular_data}, TiO, FeH, and ZrO (B. Pelz private communication). Fe ionization-excitation equilibrium is used in BACCHUS to derive effective temperature (\teff), surface gravity, iron abundance ([Fe/H]), and microturbulent velocity (\vmic). Other broadening sources (e.g., rotation, instrument resolution, macroturbulence) are modeled by a Gaussian convolution.

Stellar parameters and the convolution kernel size are derived in BACCHUS using the \texttt{param} module. We first optimize for the convolution. The we determine \teff \ by requiring a null trend,within 1$\sigma$ of the fit, in Fe I abundance versus excitation potential. We determine \logg \ by enforcing the abundances from Fe I and Fe II lines to agree within the line-by-line rms and \vmic \ by enforcing that there is no trend in iron abundance to reduced equivalent width (i.e., EW/$\lambda$). The metallicity is taken as the mean of the Fe I lines relative to the Sun. The parameter fitting uses 94 Fe I lines and 32 Fe II lines, and the details can be found in the full version of Table \ref{tab:line_selection_subset}. The reported errors in \teff, \logg, and \vmic \ are calculated through the usual propagation of error from the determination of EW of individual lines. We note that, currently, BACCHUS does not estimate the covariances between stellar parameters.

After optimizing the stellar parameters, abundances for 24 elements are measured in BACCHUS with the \texttt{abund} module. We measure abundances for the following elements: Li, C, Na, Mg, Al, Si, Ca, Sc, Ti, V, Cr, Mn, Fe, Co, Ni, Cu, Zn, Sr, Y, Zr, Ba, La, Nd, and Eu. The \texttt{abund} module synthesizes spectra at different [X/H], fixing the atmospheric model parameters to the best fitting stellar parameters. This model is generated through interpolation of nearby MARCs atmospheres; we determine the abundance for each line by minimizing the unweighted $\chi^2$ of the observed and synthesized spectrum, and we repeat this process for all selected absorption lines for that species. The line list selection is given in Table \ref{tab:line_selection_subset}. We select lines based on visual inspection of the synthesis compared with observed spectra for three stars, CM07A, CM11A, and CM25A. We select these stars because they are representative for different atmospheric parameters and rotational broadening. On top of that, for every star in our sample, selected lines must pass an internal quality check to be considered usable for that object. The internal quality check corresponds to a decision tree regarding whether the abundance output is physically plausible and constrained by the trial solutions \citep[see Section 2.2 of][for more details on the internal quality check]{BACCHUS_kh_description}. As a consequence of the internal quality check, the lines with measured abundances may vary between pairs. But we emphasize that for each pair, we adopt the same set of lines for the differential analysis.

\begin{deluxetable}{ccccccc}
\tabletypesize{\footnotesize}
\tablecaption{A portion of our line list selection. A full machine-readable version, including abundances for each star for each line, is available online. The lines used will vary between stars because of the quality checks. $\chi$ is the excitation potential, and log($A_\chi$) is the absolute abundance (before subtracting the Solar abundances) derived for this line. The solar abundances adopted are from \cite{grevesse_solar_abunds}, except where described otherwise in Section \ref{bacchus_section}. Reference keys can be matched with the full citation in Table \ref{tab:line_refs} to determine the origin of the log $gf$ values used.\label{tab:line_selection_subset}}

\tablehead{Identifier & Element & $\lambda$ & $\log gf$ &Reference Key& $\chi$ & $\log(A_X)$ \\
& & ($\mathrm{\AA}$) & (dex) & & (eV) & (dex)}
\startdata
CM00A & Na I & 5682.630 & -0.706 & GESMCHF & 2.102 & 6.053 \\
CM00A & Na I & 5688.200 & -0.404 & GESMCHF & 2.104 & 6.291 \\
CM00A & Na I & 6154.220 & -1.547 & GESMCHF & 2.102 & 6.181 \\
CM00A & Na I & 6160.740 & -1.246 & GESMCHF & 2.104 & 6.196 \\
CM00B & Na I & 5682.630 & -0.706 & GESMCHF & 2.102 & 6.103 \\
CM00B & Na I & 5688.200 & -0.404 & GESMCHF & 2.104 & 6.311 \\
CM00B & Na I & 6154.220 & -1.547 & GESMCHF & 2.102 & 6.212 \\
CM00B & Na I & 6160.740 & -1.246 & GESMCHF & 2.104 & 6.233 \\
CM01A & Na I & 4751.820 & -2.078 & GESMCHF & 2.104 & 6.558 \\
CM01A & Na I & 5682.630 & -0.706 & GESMCHF & 2.102 & 6.528 \\
\enddata
\end{deluxetable}

We study the abundance differences using both line-by-line (differential) and non-line-by-line (nLBL) ``global'' methods. Both methods provide similar conclusions; however, the differential method generally leads to better precision because it mitigates systematic errors in the spectral models. We will adopt the abundances derived with the differential method throughout this study. But for completeness, we include results from both approaches in Table \ref{tab:params_and_abund_table}. For clarity, we denote abundances for individual lines with a subscript i. For example, [X$_{\mathrm{i}}$/H] indicates an abundance of species X for line i, $\Delta$[X$_{\mathrm{i}}$/H] is the difference of species X using the abundance of line i between two stars. Species without subscripts are the overall estimate of the abundance from all corresponding spectral features.

In the differential approach, we compute $\Delta$[X$_{\mathrm{i}}$/H] for each transition observed in both stars. The abundance difference is taken as the median of these values (i.e., $\Delta$[X/H] = median($\Delta$[X$_{\mathrm{i}}$/H])). The median is used instead of the mean because it is more robust against outliers. As for the certainty estimation, we assume the statistical errors for $\Delta$[X/H], i.e., std($\Delta$[X$_{\mathrm{i}}$/H])/$\sqrt{{\rm number\; of\;lines}}$, as the standard error of the mean of the line abundance differences.  However, if a single line remains after quality cuts, we take the error as 0.1 dex. This is 1/3 the step size used between trial values in [X/H] when calculating the abundance for each line.

Abundance errors from the uncertainty in \teff, \logg, and \vmic \ are propagated according to \cite{wide_binaries_KH}. For each parameter, we perturb the best fit model and derive abundances for this perturbed model atmosphere. The difference between the abundances from the best fit and perturbed model is the error introduced from that parameter. These abundance errors are added in quadrature with the line-by-line statistical errors for $\Delta$[X/H] to determine the total error for an abundance measurement. For simplicity, and due to the BACCHUS code’s limitations, we choose to omit the covariances following \cite{wide_binaries_KH}. 

We perform the error analysis on a subset of 30 representative stars distributed evenly over the stellar parameters of our data set. The offsets for \ \teff, \logg, and \vmic \ were $\pm20$ K, $\pm0.1$ dex, and $\pm0.03$ km s$^{-1}$, respectively. The values for \teff \ and \vmic \ are the median $1\sigma$ errors reported by BACCHUS for our sample. We assume a $\Delta\,$\logg = 0.1 dex instead of median reported errors from BACCHUS (0.20 dex) because we find that BACCHUS tend to overestimate the errors of \logg \ due to some outlying Fe II lines. We have separately analyzed these spectra with Bayesian stellar parameter code LoneStar (Nelson et al in prep), which agrees with BACCHUS estimates on all quantities besides the \logg \ errors, which are found to be $\sim 0.05$ dex. We chose \logg \ errors of 0.10~dex to be conservative since BACCHUS favors larger \logg \ errors. We note however that our comparison between the close comoving pairs, far comoving pairs, and random pairs, is robust and independent of our error estimates. 

When we estimate the errors for individual observations, we adopt interpolation by inverse distance weighting (IDW) to estimate the abundance errors resulting from the uncertainties in the stellar parameters for the remainder of the data set. IDW interpolation approximates points outside the reference set by constructing a weighted average of points from the reference sets. These weights are the inverse rms distance. During the interpolation, we transform the stellar parameters onto the interval (-1,1) to weigh each equally. We tested the accuracy of the interpolation using a leave-one-out cross-validation for each element. The average error introduced was typical $< 2\%$ of the true errors calculated for that particular star. Sr and Eu were the only two elements with larger relative errors at $\lesssim 5\%$. Nonetheless, these errors are at least one order of magnitude below the total error budget for each element and are therefore negligible. 

Table \ref{tab:params_and_abund_table} summarizes the derived stellar parameters and abundance measurements and uncertainties in this study. In addition to the total abundance errors, we also report the errors from the individual errors contributed by the stellar parameter error propagation and the line-to-line scatter of $\Delta$[X/H], respectively. The stellar parameters are the main source of uncertainty. For 23 elements, the stellar parameters account for $>90\%$ of total error. For Sr, the stellar parameters accounting for $\sim 70\%$ of the total errors. Of the remaining species, Li and Eu typically only have a single line measured and are therefore subjected to the 0.1 dex error floor on the abundance scatter. Only in these rare cases, the ``line-to-line'' error is more dominant than the stellar parameters’ uncertainty. Finally, for completeness, although not shown, the nLBL abundances and errors are also provided in the full machine-readable table. 


\section{Results}
\label{Results}

\begin{splitdeluxetable*}{cccccccccccccBccccccccccccBcccccccccccc}
\tabletypesize{\scriptsize}

\tablecaption{Stellar parameters and elemental abundances derived from BACCHUS. A and B subscripts denote the two components from the comoving pairs. $\xi$ denotes the microturbulence. The electronic table contains measurements from nLBL and differential methods. All the differential measurements (e.g., $\Delta\mathrm{[Fe/H]}, \sigma_{\Delta\mathrm{[Fe/H]}}$) are measured with differential method. But we also provide individual estimates of each component with the nLBL method, indicated with the subscripts A and B. By definition, the differential analysis only applies to the differences of the two components, and there will be no individual values with the differential method. The full table also separately lists the errors from differential stellar parameters (denoted with a subscript $\theta$) and the statistical line-to-line scatter (denoted with a subscript $\mathrm{line}$), but below we only show the total errors (adding in quadrature). $\Delta\mathrm{RV}$ refers to the difference in radial velocity measured from our observations and Gaia eDR3. The large $\Delta\mathrm{RV}$ column indicates whether the system was flagged for potentially being a hierarchical triplet with an unresolved binary.\label{tab:params_and_abund_table}}

\tablehead{Component A ID & Component B ID & Identifier & Large $\Delta\mathrm{RV}$ & $\mathrm{T_{eff_{A}}}$ & $\sigma_{\mathrm{T_{eff_{A}}}}$ & $\mathrm{T_{eff_{B}}}$ & $\sigma_{\mathrm{T_{eff_{B}}}}$ & $\log g_{\mathrm{A}}$ & $\sigma_{\log g_{\mathrm{A}}}$ & $\log g_{\mathrm{B}}$ & $\sigma_{\log g_{\mathrm{B}}}$ & $\xi_{\mathrm{A}}$ & $\sigma_{\xi_{\mathrm{A}}}$ & $\xi_{\mathrm{B}}$ & $\sigma_{\xi_{\mathrm{B}}}$ & $\Delta\mathrm{[Fe/H]}$ & $\sigma_{\Delta\mathrm{[Fe/H]_{Total}}}$ & $\Delta\mathrm{[Na/H]}$ & $\sigma_{\Delta\mathrm{[Na/H]_{Total}}}$ & $\Delta\mathrm{[Mg/H]}$ & $\sigma_{\Delta\mathrm{[Mg/H]_{Total}}}$ & $\Delta\mathrm{[Al/H]}$ & $\sigma_{\Delta\mathrm{[Al/H]_{Total}}}$ & $\Delta\mathrm{[Si/H]}$ & $\sigma_{\Delta\mathrm{[Si/H]_{Total}}}$ & $\Delta\mathrm{[Ca/H]}$ & $\sigma_{\Delta\mathrm{[Ca/H]_{Total}}}$ & $\Delta\mathrm{[Sc/H]}$ & $\sigma_{\Delta\mathrm{[Sc/H]_{Total}}}$ & $\Delta\mathrm{[Ti/H]}$ & $\sigma_{\Delta\mathrm{[Ti/H]_{Total}}}$ & $\Delta\mathrm{[V/H]}$ & $\sigma_{\Delta\mathrm{[V/H]_{Total}}}$ & $\Delta\mathrm{[Cr/H]}$ & $\sigma_{\Delta\mathrm{[Cr/H]_{Total}}}$ & ... \\
&  &  &  &  (K) &  (K) &  (K) &  (K) &  (dex) &  (dex) &  (dex) &  (dex)  & (km s$^{-1}$) & (km s$^{-1}$) & (km s$^{-1}$) & (km s$^{-1}$) & (dex) & (dex) & (dex) & (dex) & (dex) & (dex) & (dex) & (dex) & (dex) & (dex) & (dex) & (dex) & (dex) & (dex) & (dex) & (dex) & (dex) & (dex) & (dex) & (dex) &  ...}
\startdata
6160638833832941312 & 3471119180522314624 & CM00 & False & 5809 & 9 & 5934 & 65 & 4.53 & 0.30 & 4.65 & 0.41 & 1.22 & 0.04 & 1.31 & 0.05 & 0.04 & 0.02 & 0.03 & 0.02 & 0.02 & 0.02 & -0.02 & 0.02 & 0.01 & 0.02 & -0.01 & 0.05 & 0.04 & 0.06 & 0.07 & 0.03 & 0.09 & 0.04 & 0.05 & 0.03 & ... \\
3613454637229633664 & 3613454637229634176 & CM01 & False & 6058 & 25 & 6002 & 56 & 4.29 & 0.16 & 4.14 & 0.28 & 1.30 & 0.04 & 1.33 & 0.04 & 0.02 & 0.02 & 0.05 & 0.02 & 0.03 & 0.02 & 0.03 & 0.01 & 0.01 & 0.02 & 0.01 & 0.03 & 0.02 & 0.05 & 0.02 & 0.03 & 0.01 & 0.03 & 0.06 & 0.03 & ... \\
3639520621950395904 & 3639520621950395776 & CM02 & False & 5962 & 11 & 5966 & 11 & 4.38 & 0.22 & 4.41 & 0.23 & 1.06 & 0.04 & 1.05 & 0.04 & 0.00 & 0.02 & 0.00 & 0.02 & 0.01 & 0.02 & -0.02 & 0.01 & 0.01 & 0.02 & -0.02 & 0.04 & 0.01 & 0.06 & -0.00 & 0.03 & -0.01 & 0.04 & 0.01 & 0.03 & ... \\
5897704951769034368 & 5897785667103983616 & CM03 & False & 6036 & 37 & 6121 & 28 & 4.34 & 0.22 & 4.28 & 0.21 & 1.31 & 0.04 & 1.34 & 0.04 & -0.08 & 0.02 & -0.11 & 0.02 & -0.06 & 0.02 & -0.12 & 0.02 & -0.08 & 0.02 & -0.06 & 0.03 & -0.10 & 0.05 & -0.07 & 0.03 & -0.09 & 0.04 & -0.09 & 0.03 & ... \\
6224633983987510528 & 6224633983987511552 & CM04 & False & 5763 & 26 & 5822 & 11 & 4.51 & 0.14 & 4.50 & 0.20 & 0.90 & 0.04 & 0.99 & 0.04 & -0.01 & 0.02 & -0.00 & 0.03 & 0.01 & 0.01 & 0.00 & 0.01 & -0.03 & 0.03 & -0.00 & 0.05 & -0.04 & 0.07 & 0.00 & 0.03 & 0.05 & 0.04 & 0.03 & 0.04 & ... \\
\enddata
\end{splitdeluxetable*}

\begin{figure*}
    \centering
    \includegraphics[width=2\columnwidth]{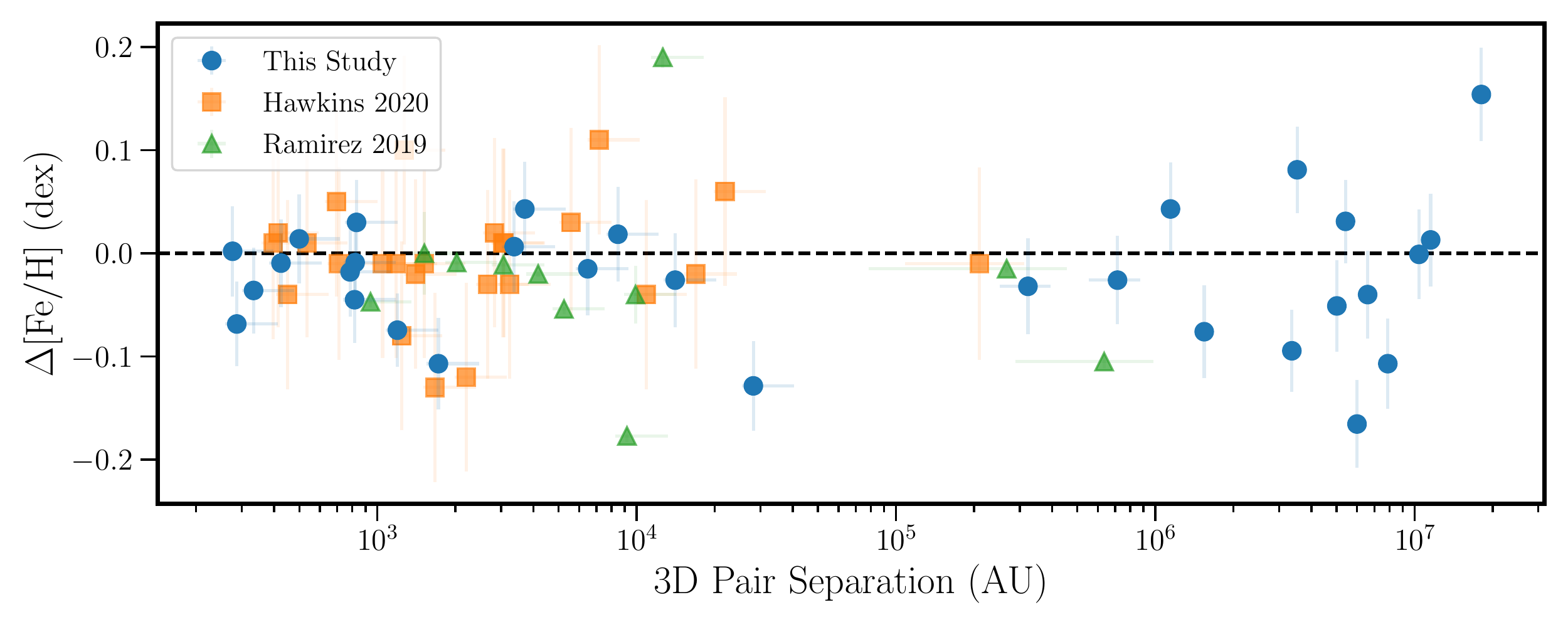}
    \caption{A comparison of the comoving pairs (blue) with similar works on wide binaries. Separation errors for pairs with separations $> 10^6 \mathrm{\ AU}$ are essentially negligible and smaller than the symbol size. Our close comoving pairs (i.e., wide binaries) have a similar degree of chemical homogeneity to the literature values. Far comoving pairs (pairs with separations $> 2\times10^5 \mathrm{\ AU}$) are also chemically homogeneous in [Fe/H], with a scatter of $<0.1\,{\rm dex}$, albeit being slightly less homogeneous than close comoving pairs.}
    \label{fig:delta_FeH_vs_sep}
\end{figure*}

As discussed in the introduction, this paper’s main goal is to assess whether or not the comoving pairs in this study, especially for the far comoving pairs, are conatal systems. Since conatalitly cannot be assessed directly outside star-forming regions, we resort to studying the chemical homogeneity (i.e., $\Delta\mathrm{[X/H]}$). Figure \ref{fig:delta_FeH_vs_sep} displays the difference in metallicity between comoving pairs (blue) as a function of pair separation. Although not shown, similar trends are seen for the majority of the elements studied.

For comparison, we also include the wide binaries results from \cite{Ramirez2019}, and \cite{wide_binaries_KH}. We note that \cite{Ramirez2019} includes results for 12 systems, 11 of which are from other studies \citep[i.e.,][]{Ramirez_2010,Liu_2014, Mack_2014, Ramirez_2014a, Tucci_Maia_2014, Teske_2016, Saffe_2016, Mack_2016, Saffe_2017, Reggiani_2018, Oh2018}. In the remainder of the paper, we will exclude the system from \cite{Mack_2014} from comparisons because it appears to be a hierarchical triplet with an unresolved binary, and use \cite{Ramirez2019} as a shorthand for the remaining 11 systems. We note that since the sample \citet{Ramirez2019} is a heterogeneous sample, the comparison with \citet{wide_binaries_KH} might be more relevant. Furthermore, the samples in \cite{wide_binaries_KH} have spectra with spectra and were analyzed in the same way as this study. We also compare our data to \cite{Andrews2019}; however, they did not provide individual measurements, so our comparison with this work is limited to summary statistics. 

As previously discussed, we split our comoving sample into the close (i.e., primarily wide binaries) and far comoving pairs to compare with these works. Close comoving pairs are defined as those having separations below 1 pc, i.e. $2\times10^5$ AU, (with 17 pairs in this study), and far comoving pairs have separations above 1 pc (14 pairs). For the close pairs, we use the geometric prior when calculating the separations (see Section \ref{separations_velocities_method1}) because the parallax error dominates. 
Our close comoving pairs sample (i.e., wide binaries) show a similar degree of chemical homogeneity as other studies on wide binaries with a dispersion of 0.05 dex in $\Delta$[Fe/H], which is expected. Compared to the close comoving pairs, we find that the far comoving pairs are less chemically homogeneous, with a $\Delta$[Fe/H] dispersion of 0.08 dex; however, the far comoving pairs are substantially more homogeneous than the random pairs (0.23 dex), which we will further explore in the next section. Table \ref{tab:med_scatter} summarizes the scatter in $\Delta$[Fe/H] comparing our data with prior works. 

\begin{table}
    \centering
    \begin{tabular}{l|c|c}
    & median $|\Delta[\mathrm{Fe/H}]|$ (dex) & std $\Delta[\mathrm{Fe/H}]$ (dex) \\
    \hline
    \hspace{-0.2cm}{\bf Wide Binaries} & & \\
        \cite{wide_binaries_KH} & 0.02 & 0.05 \\
        \cite{Ramirez2019} & 0.04 & 0.09\\
        \cite{Andrews2019} & - & 0.04\\
         close comoving pairs & 0.03 & 0.05\\[0.1cm]
   \hline
       \hspace{-0.2cm}{\bf Unbound pairs} & & \\
         far comoving pairs & 0.05 & 0.08 \\[0.1cm]
    \hline
    	 random pairs & 0.16 & 0.23 \\
    \end{tabular}
    \caption{A comparison of the $\Delta\mathrm{[Fe/H]}$ scatter for our study and three other studies on wide binaries. \cite{Andrews2019} only provides standard deviation without giving the individual measurements, which prohibits us from calculating the median of the absolute difference. \cite{Andrews2019} provides two values for the scatter in $\Delta[\mathrm{Fe/H}]$, one for the entire sample (0.08 dex) and the other (0.04 dex) for wide binaries with $\Delta$\teff$< 100$ K. We only state the latter as it is more relevant to this study. We caution that \cite{Ramirez2019} is a heterogeneous sample of several studies. As such, it could incur a large scatter due to different systematics from these studies, the comparison to the other studies might be less direct.} 
    \label{tab:med_scatter}
\end{table}

\subsection{Chemical Homogeneity of Comoving Pairs}
\label{xh_section}

\begin{figure}
    \centering
    \includegraphics[width=1\columnwidth]{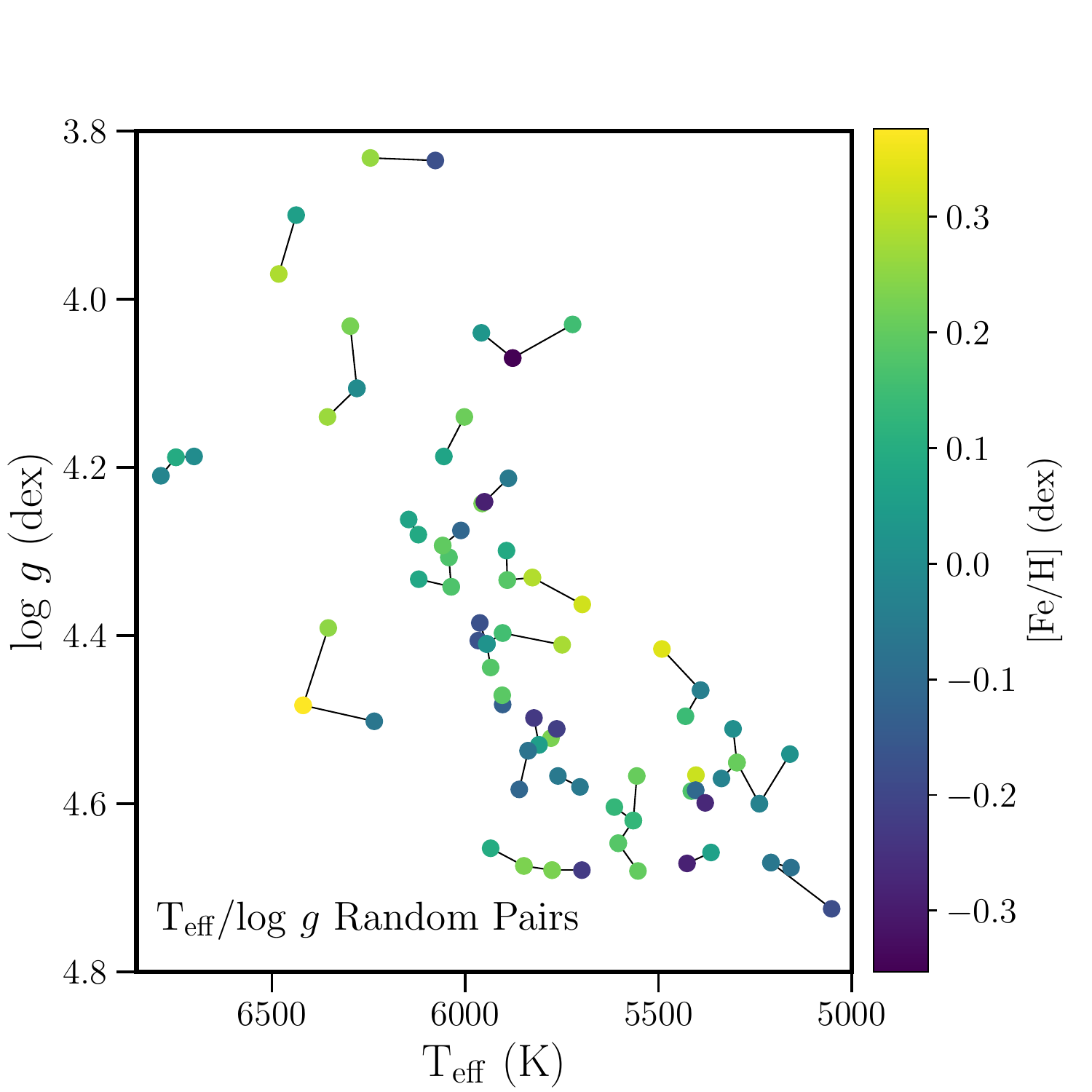}
    \caption{The stellar parameters of the simulated random field pairs in this study. To simulate unrelated pairs of field stars and compare that to the comoving pairs, we assign a partner based on the nearest star in stellar parameter space, which is not its comoving counterpart. The black lines connect paired stars by minimizing the distance in \teff \  and \logg. A single star may be paired with multiple partners. This was allowed because we want to prevent pathological assignments where two stars are paired over large distances because all intermediates are unavailable, artificially inflating systematic uncertainties.}
    \label{fig:random_pairs_diagram}
\end{figure}

To contextualize how homogeneous our comoving population is, we simulate pairs of random field stars. In a similar fashion to \cite{wide_binaries_KH}, we create a set of random pairs by assigning every star to another star, which is not its comoving partner. Since the comoving stars are selected to be stellar twins, for better comparison, we create random pairs which minimize the difference in \teff \ and \logg \ (labeled \teff/\logg \ pairs) to mitigate systematic effects. Since \teff \ and \logg \ have different units, when minimizing the difference, we divide both quantities by the standard deviation of the pairwise difference for all non-comoving pairs of stars in our sample. The random pairs are shown in Figure \ref{fig:random_pairs_diagram}. Across all elements, the average dispersion in $\Delta$[X/H] for the random pairs is 0.24 dex, significantly larger than the dispersion of both the close comoving pairs (0.05 dex) and the far comoving pairs (0.10 dex).

\begin{figure*}
    \centering
    \includegraphics[width=2\columnwidth]{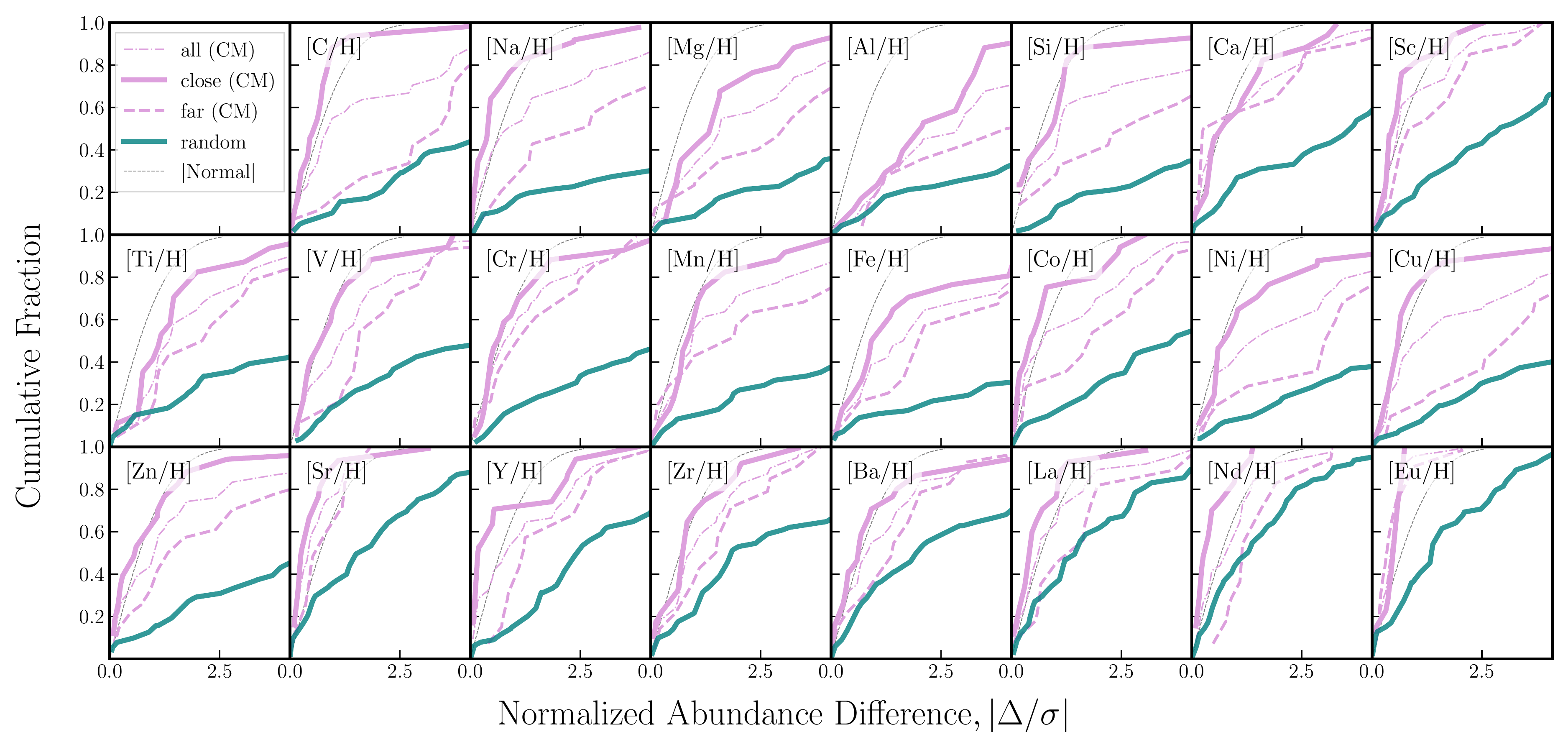}
    \caption{The distribution of difference in elemental abundances normalized by their uncertainties. The comoving pairs are divided into three groups (the entire comoving sample; violet thin dashed line), close (violet thick solid line), and far (violet dashed line, medium width). The thin black line indicates an expected distribution if the scatter in abundance differences between comoving pairs were explained by the measurement uncertainties. The solid teal line indicates the distribution for the random pairs. The close comoving pairs are more chemically homogeneous than the random groups and generally approaches a distribution consistent with the observed spread being only a result of measurement errors.  The far comoving pairs are less chemically homogeneous than the close comoving pairs. However, they are generally more chemically homogeneous than the random pairs, indicating a mixed population of conatal pairs and chance alignments.}
    \label{fig:delta_XH}
\end{figure*}


To better illustrate the difference in chemical homogeneity between these groups, similar to \cite{Andrews2019}, we visualize the homogeneity of our sample using $|\Delta\mathrm{[X/H]}|/\sigma_{\Delta\mathrm{[X/H}]}$, which will be abbreviated as $\Delta/\sigma$. The $\Delta$ here is the difference in elemental abundances between the two components, and the $\sigma$ is the quadrature sum of the uncertainties from the two components. Figure \ref{fig:delta_XH} shows the cumulative distribution function (CDF) of $\Delta$[X/H] normalized by the measurement error for the comoving pairs (violet/solid and dashed) and  random pairs (teal/solid). For reference, each panel includes the CDF of $\Delta/\sigma$ drawn from the unit Gaussian -- i.e., what a chemical homogeneous would look like at the level of our current measurement uncertainties (assuming that our uncertainty estimations are accurate). 

There is a clear difference in the random pairs’ distribution and comoving pairs, both close and far, largely attributed to the larger chemical dispersion seen in the random pairs. The difference in the CDF between the comoving and random pairs also varies between elements. This variation is primarily caused by the difference in our measurement precision for those elements. On the one hand, for better-measured elements, such as Fe, Mg, Al, and Si, the comoving pairs exhibit a more distinct chemical homogeneity than the random pairs. On the other hand, unsurprisingly, for less well-measured elements, the difference in chemical homogeneity is less distinct, as shown in the CDFs for La, Nd, and Eu. We do not include Li in this plot because its surface abundance has a strong dependence on effective temperature and is therefore not a useful indicator for conatality \citep[see][]{gray_book, Ramirez2012_Li, wide_binaries_KH}.

For most elements, the close comoving pairs are consistent with a chemically homogeneous distribution at our measurement precision, with the CDF for 16 elements exceeding 96\% by $\Delta/\sigma = 4$. In contrast, the random pairs are less homogeneous, with the CDF for 13 elements $\lesssim50\%$ at $\Delta/\sigma = 4$. 


As already seen in Fig.~\ref{fig:delta_FeH_vs_sep}, the far comoving pairs are more heterogeneous than the close comoving pairs for most elements, albeit still being more homogeneous than the random pairs. The slightly larger chemical inhomogeneity is unsurprising because we expect some of the far comoving pairs might be contaminated by chance alignment pairs \citep[e.g.,][]{Kamdar2018}, which we will quantify in the next section. 

In this study, we focus on the abundance [X/H] instead of the abundance ratio [X/Fe]. However, all elements correlate with iron in [X/H]. Therefore, the difference between random and close comoving pairs in $\Delta/\sigma$ for [X/H] may only reflect a metallicity difference.  While [X/H] (or metallicity alone) are tell-tale signs of conatality (or at least co-eval), studying the abundance ratio [X/Fe] might, in principle, constitute a more stringent test and might reveal subtle physics. Although not shown, we tried to examine $\Delta/\sigma$ for [X/Fe]\footnote{For this test, we minimized the difference in \teff\ and [Fe/H] when simulating the random field pairs to find pairs of stars with the same metallicity.}. We find little difference between the comoving and random pairs of stars in the [X/Fe] space. This is not unexpected because the residual variance in [X/Fe] is much more subtle than in [X/H]. For example, \cite{Ting2021} shows that when one subtracts the mean chemical track, the residual correlation in [X/Fe] is small, with signal $\lesssim 0.01-0.02$ dex. Therefore, to see such a signal, we would need to measure [X/Fe] better than $\sim 0.01\,$dex. With the current pipeline adopted in this study, we found that achieving 0.01 dex remains challenging. Therefore, we only focus on [X/H]. Future studies with even higher quality spectra and/or better analysis will be needed to shed light on these subtle variations.

\subsection{Conatal Fraction for Separations $>2\times10^5\,$AU}
\label{co-natal_far}

The close comoving pairs (wide binaries) at separations below $2\times10^5\,{\rm AU}$ are commonly believed to be conatal as chance capture is unlikely. However, beyond such separation, the conatality for the far comoving pairs remains unexplored. Armed with our high-resolution spectroscopic data from these pairs, in this section, we will constrain how conatal the far comoving pairs with separations $>2\times10^5\,$AU are relative to the close comoving pairs. Showing the conatality of far-comoving pairs can have far-reaching consequences; if these comoving pairs are chemically homogeneous and conatal, this would significantly expand the sample of calibrators beyond open clusters and wide binaries.

As we have seen in the previous section, these far comoving pairs are more heterogeneous than the close comoving pairs/wide binaries but more chemically homogeneous than the random pairs. One explanation for the slightly larger chemical inhomogeneity is that the far comoving pairs are a mixture of conatal pairs and chance alignments. The interlopers create a tails of large $\Delta [{\rm X/H}]$. If that is the case, we should be able to constrain the conatality fraction through these far-comoving pairs’ chemical homogeneity, which is what we will attempt next.

We only model $\Delta\,$[Fe/H] because iron has the most precise measurements of any element in our sample, and, as we have argued, most elemental abundances only trace [Fe/H] at the current precision. The basic idea is that, if the $\Delta\,$[Fe/H] of the far comoving pairs are contributed by the two populations -- a conatal population that resembles the wide binaries population and the random pairs interlopers, we would expect the distribution of  $\Delta\,$[Fe/H] would also be a mixture of the two. More specifically, we model the PDF of $|\Delta\mathrm{[Fe/H]}|/\sigma_{\Delta\mathrm{[Fe/H}]}$ (or in short, $\Delta/\sigma$, in the following) for the far comoving pairs as a mixture model using the weighted average of the close comoving pairs' $\Delta/\sigma$ PDF and the random pairs' $\Delta/\sigma$ PDF. The best-estimated weight signifies the fraction of pairs consistent with the close comoving pairs among the far comoving populations, hence its conatal fraction. Due to our small sample size ($\sim 10$ pairs), we expect considerable uncertainty for estimating the conatal fraction because of the sampling noise. Therefore, to properly model the sampling noise, we also estimate the uncertainty of the estimated conatal fraction by bootstrapping the sample.

Operatively, first, we sample, with replacement, the data distribution of $\Delta/\sigma$ for the close, far, and random pairs. The distributions of points drawn from the close comoving pairs and random pairs are then separately approximated as normal distributions through maximum likelihood estimation (MLE). Subsequently, we fit for the weight of the far comoving pairs distribution, also through MLE, assuming that the $\Delta/\sigma$ distribution of the far comoving pairs is a weighted mixture of the ones determined for the close moving pairs and the random pairs. The bootstrapping process is repeated 1000 times. The bootstrapping produces a distribution of weights. The median and standard deviation are taken as the best estimate and uncertainty of the conatal fraction, respectively.

\begin{figure}
    \centering
    \includegraphics[width=1\columnwidth]{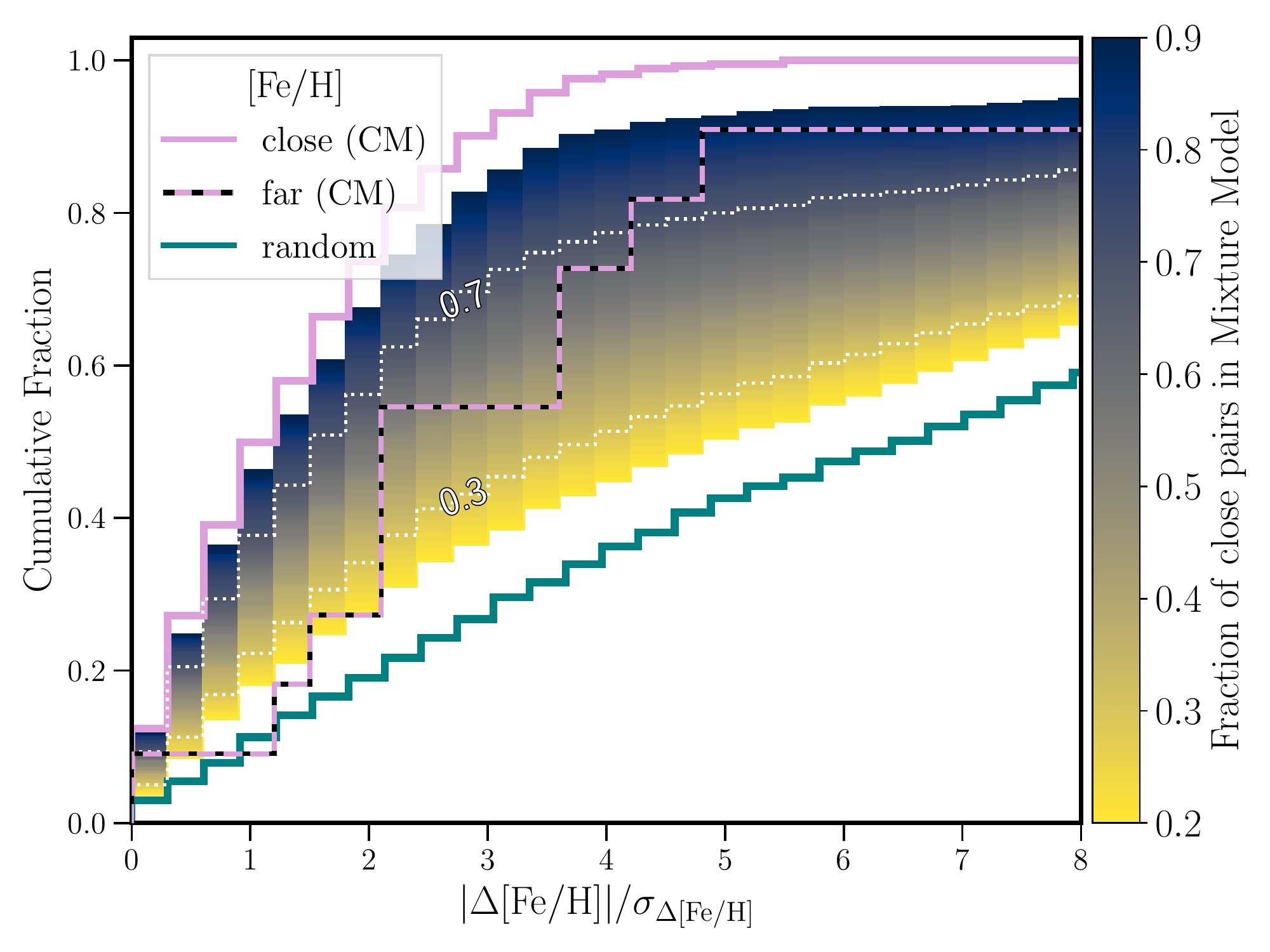}
    \caption{ The normalized chemical homogeneity CDFs for different populations in this study. The CDF from the close comoving pairs/wide binaries is plotted in the bold violet line and the random field pairs in the bold teal line. The colored region represents a mixture model of the close and random pairs using different mixing ratios. The far comoving pairs (violet/black dashed) with separations of $2\times10^5-10^7\,$AU is best modeled with a mixture with a 73\%/27\% ratio of the close comoving pairs versus the random pairs. If all close comoving pairs are conatal, the chemistry of stars reveals that about 73\% of the far comoving pairs are conatal.}
    \label{fig:cdf_mixture}
\end{figure}

In Figure \ref{fig:cdf_mixture} we demonstrate the nominal fit without the replacement. For this estimate, we only consider comoving pairs with separations $2\times10^5-10^7\,$AU, excluding the three pairs with separations $>10^7\,$AU, leaving us with 11 pairs (out of 14 pairs). As we will see in the following, these three pairs are clearly interlopers (based on predictions from simulations). The $\Delta/\sigma$ CDF of the close moving pairs is shown in violet, the far comoving pairs in violet/black, and the random pairs in teal. A continuum of mixture model CDFs is displayed in the background as a blue/yellow gradient ranging from close:random ratios of 1:4 to 9:1, or equivalently, a conatal fraction 20\%-90\%. Besides, we highlight some fractions with white dotted contours to guide the reader. We find the conatal fraction for the 11 far comoving pairs with separations $2\times10^5-10^7\,$AU to be $73\pm22\%$. In short, our result implies that $\sim 75\%$ of the far comoving pairs are conatal based on their metallicity measurements.

\begin{figure*}
    \centering
    \includegraphics[width=2\columnwidth]{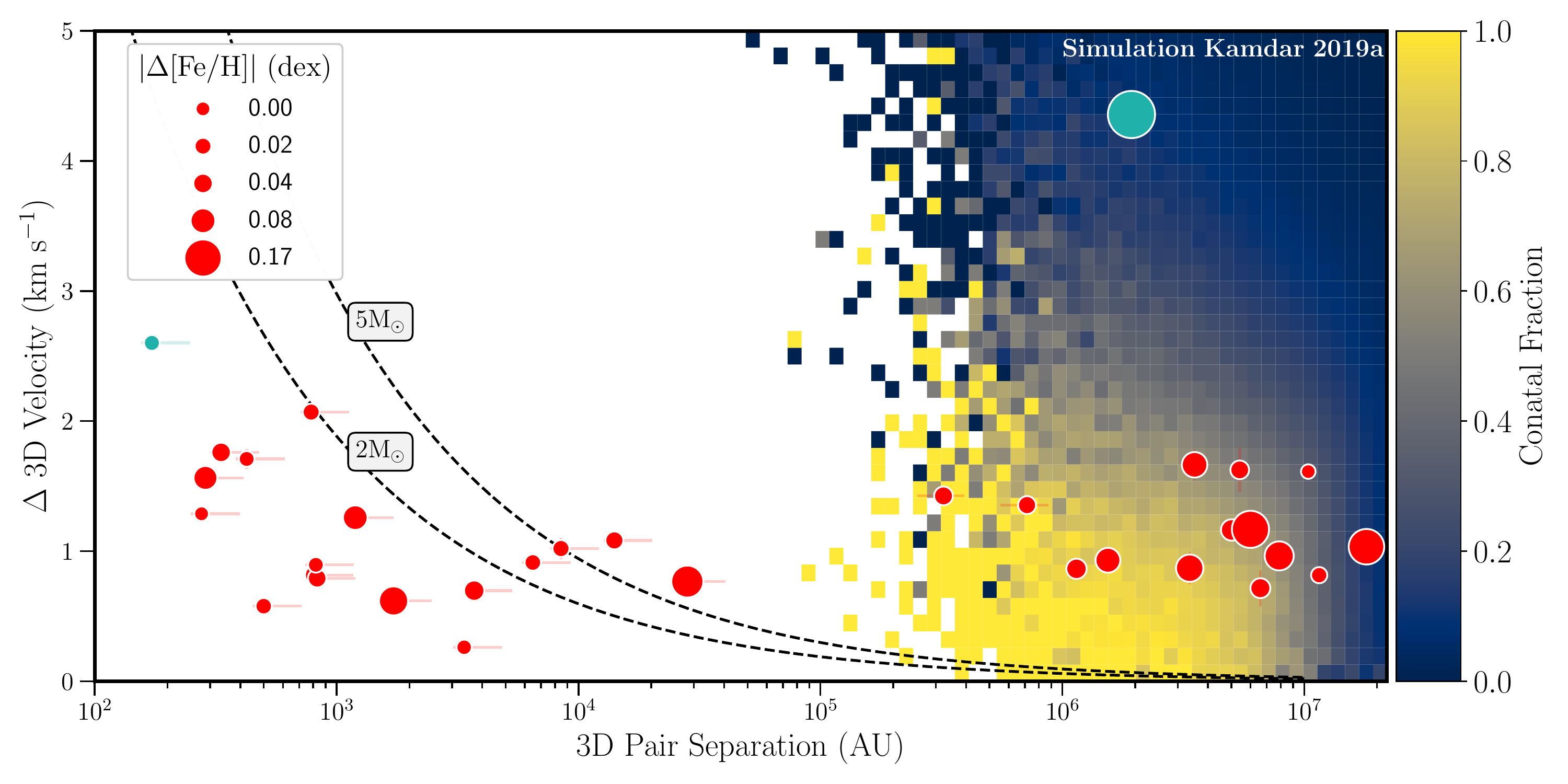}
    \caption{A comparison of our main sample (red circles) with the theoretical predictions from simulation \citep{Kamdar2018}. The symbol sizes are scaled relative to the two stars’ metallicity differences. A smaller metallicity difference indicates a larger likelihood of being a conatal pair. The background shows the expected conatal fraction according to the simulation.  The simulation does not include wide binaries and therefore does not extend to separations below $10^5\,$AU. The black dashed contours on the left indicate the maximum velocity difference a bound system could have with a given total mass, assuming equal-mass companions and no unseen tertiaries. Our data show excellent agreement with the simulation, with smaller metallicity differences within the predicted locus of conatal stars and more considerable metallicity differences outside the locus. We include two systems (bluegreen) that have velocity difference $> 2\,$ km \ s$^{-1}$ as references. The first is a system with a velocity difference of $2.5$ km \ s$^{-1}$ and a separation of $2 \times 10^2\,$AU, this is well within the parameter space for a binary system and is likely conatal. The second has a larger spatial separation ($2\times 10^6\,$AU) and a velocity difference greater than 4 km \ s$^{-1}$. The simulation predicts that such a pair is likely a chance alignment. The abundance differences of these two pairs concur with these expectations.}
    \label{fig:simulation_comp}
\end{figure*}

\section{Discussion}
\label{discussion}

Conatal stars are the central hallmarks in modern-day Galactic Archaeology. Absolute standards are hard to come by \citep{benchmark_metal, benchmark_library, benchmark_temperature_grav}, but if we know the two stars are conatal, the two components serve as each other references. This is why open clusters and wide binaries have always been the golden calibrators to refine stellar models, to study exoplanets, gas mixing, and to calibrate surveys. More recently, the high precision astrometry measurements from Gaia now allow us to identify wide binaries at high fidelity, which further propels the study of wide binaries. However, most studies of wide binaries \citep[e.g.,][]{elbadry_wb_edr3_preprint} often assume a dichotomy between wide binaries and chance alignments, and this picture is clearly too simplistic. Simulations \citep{Kamdar2018,kamdar2019b} have suggested comoving stars from disrupted star clusters could have a high probability of being conatal out to separations of $\mathcal{O}(10^6)\,$AU. If this is correct, unbound comoving stars could be used in similar ways as wide binaries and open clusters. Based on modeling of Gaia data by Ting et al (in prep), we expect to find roughly 3300 conatal pairs with 3D separations $\lesssim 2\times10^6\,$ AU (10 pc) and $\Delta\mathrm{v_{3D}} < 2$ km s$^{-1}$.

In this study, we perform the first homogeneous study of comoving pairs that span five orders of magnitude in separation. A homogeneous sample allows us to study the chemical homogeneity of the comoving stars with different separations consistently. We found that the close comoving pairs with separations $< 2\times10^5\,{\rm AU}$ have a $\Delta$[Fe/H] scatter of $0.05\,$dex. These values are comparable to the typical values seen in open clusters and other wide binaries studies. In particular, we opt to perform the same analysis as \citet{wide_binaries_KH}. With similar resolution and SNR, \citet{wide_binaries_KH} showed a dispersion of 0.05 dex for wide binaries, largely consistent with this study. On top of that, we show that the abundance differences for almost all elements are consistent with the measurement uncertainties, further validating that close comoving pairs/wide binaries are conatal. We note that some of the $\Delta\mathrm{[Fe/H]}$ scatter seen in the systems from \cite{Ramirez2019} could be a consequence of a selection effect. Many of those systems were published in the context of planet engulfment signatures. Consequently, the systems which show the engulfment signals \citep[e.g., Kronos and Krios][]{Oh2018}, will also have larger metal differences. Since exoplanet accretion changes the surface abundances, these systems may not be representative of the true chemical homogeneity of conatal systems.


More importantly, our study expanded on previous works by also examining comoving pairs with separation $> 10^6\,{\rm AU}$. The key result of our study is that comoving pairs at these larger separations are still significantly more chemical homogeneous than the random pairs, albeit with a slightly larger dispersion in chemistry, with a scatter  $\Delta\mathrm{[Fe/H]}$ of $0.09$ dex. The random pairs have a scatter of 0.23 dex. If we assume that the far comoving pairs are comprised of both the conatal stars and chance alignments, we estimate that about $73\pm22\%$ of pairs with separations $2\times10^5-10^7\,{\rm AU}$ are conatal.

A caveat with this estimation is that, in this study, we select stellar twins with a median $|\Delta\,$\teff$|$ of 105 K. The selection of stellar twins enables the mitigation of any potential systematics in terms of the spectral modeling by performing a line-by-line differential study. Conveniently, from their construction, the simulated random pairs have a smaller difference in stellar parameters (with a median $|\Delta\,$\teff$|$ of 47 K), so we would expect these systematic effects to be similar or weaker in the random population. However, for completeness, we investigated whether or not the more significant dispersion in metallicity difference is real or simply due to systematic errors when deriving abundances from stars that are more different in stellar parameters.

To dissect that, we looked into any potential biases in $\Delta\mathrm{[Fe/H]}$ as a function of the difference in stellar parameters. We found weak trend between $\Delta\mathrm{[Fe/H]}$ and $\Delta\mathrm{T_{eff}}, \Delta\log g,$ and $\Delta\mathrm{v_{micro}}$. For example, $\Delta\mathrm{[Fe/H]}$ shows a positive correlation with $\Delta\mathrm{T_{eff}}$, with a gradient of $2.3\times10^{-4} \mathrm{dex \ K^{-1}}$. We attempted to account for that by fitting a multivariate linear regression between $\Delta\mathrm{[Fe/H]}$ as a function of $\Delta\mathrm{T_{eff}}, \Delta\log g,$ and $\Delta\mathrm{v_{micro}}$. We found that this process typically leads to a correction of $0.002 \, \mathrm{dex}$ in $\Delta\mathrm{[Fe/H]}$, which is negligible for our study. We conclude that it is unlikely the differences in chemistry between the comoving and random pairs are attributable to this correlation. We opted not to remove these correlations because it is hard to determine the exact causal direction of this correlation.

Recall that, the observations in this study are largely motivated by the simulation from \cite{Kamdar2018}, in which the authors argued that far comoving pairs are also conatal. So, how does our study compare with the simulation? In Figure~\ref{fig:simulation_comp}, we compare our data with the \cite{Kamdar2018} simulation. Our comoving pairs are represented as red circles, and the symbol size shows the metallicity difference of the two stars for individual pairs. The black dashed lines indicate the escape velocity of bound systems, assuming equal-mass companions, and with the total masses of $2 {\rm M}_\odot$ and $5 {\rm M}_\odot$ respectively. The simulation does not extend below $10^5\,$AU because the simulations did not attempt to model wide binaries (only disrupted star clusters). 

Our results exhibit excellent agreement with the simulations; our sample shows a high degree of chemical homogeneity in the regime where the simulations predict to have a high conatality rate (the region in yellow). In the phase space regime where chance alignments should dominate (in blue background), our observations also show the largest metallicity differences (largest symbol sizes). Also more quantitatively, the simulation predicts that $\sim 80\%$ of comoving pairs are conatal provided $\Delta\mathrm{v_{3D}} < 2$ km s$^{-1}$ and the pair separation is between $2\times10^5-10^7\,$AU . Our conatality fraction estimate of $73\pm 22\%$ for separations is well aligned with the simulation's predictions. 

On top of that, besides our (31 pairs of the main sample), as discussed in the target selection, we also observed two pairs with large $\Delta\mathrm{v_{3D}}$ ($> 2\,$km/s) as the control sample. One of the two pairs has a spatial separation of $2 \times 10^2\,$AU. Even though they are not within $\Delta\mathrm{v_{3D}} < 2$ km s$^{-1}$, this pair is clearly a bound binary, and therefore is conatal. The other pair has a large separation ($2 \times 10^6\,$AU) {\em and} a large velocity difference ($4\,$km s$^{-1}$). The simulation predicted this pair is firmly a chance alignment pair as disrupted star cluster members are unlikely to exhibit such a large velocity difference. As shown in Fig.~\ref{fig:simulation_comp}, our abundance measurement indeed concurs with this picture, the former has a metallicity difference that is typical of that of wide binaries, and the latter has a dispersion of 0.24 dex, consistent with what we expect from a random pair. With this in mind, we caution that all the results presented in this study, including the conatal fraction, only strictly applies to stars with $\Delta\mathrm{v_{3D}} < 2\,$km s$^{-1}$.

While the agreement is encouraging, due to the small sample size, unfortunately, we could only settle on two separation bins -- those with separations $< 2\times10^5\,{\rm AU}$ and those with $2\times10^5-10^7\,{\rm AU}$. It is clear from the simulation that there is no sharp transition between the conatal to random pairs. But instead, the regime of purely conatal pairs transitions gradually toward the regime dominated by chance alignments. The transition depends critically on many Galactic properties, including how stars form and disperse and the cluster mass function. On top of that, as explored in \cite{Kamdar2018}, the number density of comoving stars provide tell-tale signs on the number density of Galactic perturbers, such as giant molecular clouds in the Milky Way disk, as they disrupt comoving pairs. In short, understanding this transition with critical and refining this boundary region with a larger sample will no doubt shed insights into the formation and dispersion of star clusters and other substructures in the Milky Way.

The exact value of the division between close pairs (i.e., wide binaries) and far pairs (presumed unbound) is somewhat arbitrary. Bound pairs are not expected to exist at separations beyond 1-2 pc, because at these separations the Galactic tidal field is stronger than the internal acceleration within a binary \citep[e.g.][]{Jiang_wb_tidal_radius}. At separations of order 1 pc, there will always be some ambiguity between still-bound and recently-dissolved pairs, and the reliability with which bound pairs can be identified will depend on the precision of the astrometry \citep[e.g.][]{elbadry_wb_edr3_preprint}. A possible alternative to adopting a strict threshold at ${\rm S_{eDR3}} = 2\times 10^5\,\rm AU$ would be to only consider as binaries pairs that are members of a predetermined wide binary catalog, such as the one produced by \cite{elbadry_wb_edr3_preprint}. The disadvantage of this approach is that such catalogs are generally not 100\% complete. Increasingly precise astrometry form future {\it Gaia} data releases will make it possible to classify pairs with separations of order 1 pc more reliably. For now, we note that because of our small sample size, our estimates of the conatal fraction are considerable and are dominated by small number statistics. Small changes in the adopted boundary between close and far pairs thus lead to changes in our inferred conatal fraction that are comfortably within our reported uncertainties. For completeness, we provide a summary of the effects of choice of separations for the far pairs in Table \ref{tab:conatal_frac_table}. 


\begin{table}[]
    \centering
    \begin{tabular}{c|c}
        Threshold (AU) & Conatal Fraction \\
        \hline
        $10^5$ & $52\pm26\%$\\
        $2\times 10^5$ & $73\pm22\%$\\
        $10^6$ & $61\pm31\%$\\
       
    \end{tabular}
    \caption{This table details the effects of using different thresholds for the close and far pairs. We always assume an upper separation threshold of $10^7\,$AU for the far comoving pairs.}
    \label{tab:conatal_frac_table}
\end{table}


Finally, in this study, we assume that the difference in chemical homogeneity for the far comoving pairs compared to the close conatal pairs stems from random pairs interlopers. However, another potential source for these differences in chemical homogeneity could be from the ISM being less mixed at the largest scale. At the $\mathcal{O}(10^7)\,$AU scale, ISM can have abundance scatter of up to 0.3 dex \citep[e.g.][]{Sanders2012}. If stars formed in a poorly mixed cloud, they might inherit this scatter. We deem such a scenario unlikely because hydrodynamic simulations have found that, in a single star-forming region, even small amounts of turbulence can cause a factor of five reductions in abundance to scatter in stars compared to the progenitor ISM \citep{Feng_2014}. And this chemical reduction scatter applies to scales large as $\mathcal{O}(10^6)-\mathcal{O}(10^7)\,$AU, the maximum scale probed in this study. 

That said, if the stars formed in a filamentary structure at different star-forming regions, the ISM might be less well mixed. For example, \cite{Hawkins_stream} argue Pisces-Eridanus stellar stream could have been produced from a stellar filament. They find a metallicity range of $\sim0.2$ dex for the stream, which is consistent with our metallicity difference in the far comoving pairs. If the chemical inhomogeneity is due to the ISM physics instead of random interlopers, our conatality fraction estimated would be a conservative limit. The far comoving pairs would have an even higher conatal fraction than what was is inferred here.

\section{Summary}
\label{summary}

We obtained a homogeneous sample of high-resolution ($R\simeq 40000$), high SNR ($\sim 150$ per pixel) observations of 62 FG-type stars residing in 31 comoving pairs with separations that span five orders of magnitude in separation, between $10^2\,$AU to $10^7\,$AU. Previous studies mostly restricted to wide binaries with separation $< 10^6\,$AU, and we investigate if the (unbound) comoving stars with separations $=2\times10^5-10^7\,$AU are conatal. We measured the stellar atmospheric parameters and chemical abundances for 24 species, including covering the different nucleosynthetic pathways. The derived elemental abundances: Li, C, Na, Mg, Al, Si, Ca, Sc, Ti, V, Cr, Mn, Fe, Co, Ni, Cu, Zn, Sr, Y, Zr, Ba, La, Nd, and Eu. 

We separate our sample into the classical wide binaries regime (i.e., close comoving pairs with separations $< 2\times10^5\,$AU), the far comoving pairs (separations $> 2\times10^5\,$AU), as well as the random field pairs created through random pairings of our observations. We find the wide binaries are significantly more chemically homogeneous than field stars in [X/H] (see Figure \ref{fig:delta_XH}) with a typical scatter in $\Delta$[Fe/H] of 0.05$\,$dex. Our results agree with previous works on wide binaries and comoving pairs \citep[see, e.g.,][]{Andrews2019, wide_binaries_KH}. For most elements in the close comoving pairs, the dispersion in [X/H] is consistent with the measurement errors, implying that the close comoving/wide binaries pairs are chemically homogeneous at our measurement precision.

We demonstrate that the far comoving pairs with separations of $2\times10^5-10^7\,$AU also exhibit substantial chemical homogeneity ($0.09\,$ dex in the [Fe/H] scatter) compared to random pairs (0.23 dex). Nonetheless, the far comoving pairs less homogeneous than the close comoving/wide binaries population. If we assume that the far comoving pairs comprise of a mixture of the conatal and random chance alignments population, modeling the distribution of $\Delta$[Fe/H] as a mixture of these two populations implies that about $73\pm22\%$ of the unbound comoving stars with separations $2\times10^5-10^7\,$AU and $\Delta\mathrm{v_{3D}} < 2$ km s$^{-1}$ are conatal. This conatality fraction is in excellent agreement with the predictions from simulations from \cite{Kamdar2018}. 

Our study implies that most comoving stars are conatal, even though they are well-separated. As well-separated pairs of stars are more common than wide binaries, this new population of ``clusters of two'' enables many windows for studies that we have thus far restricted to the wide binaries and open clusters. Harnessing these vastly abundant comoving pairs will have broad applications, ranging from calibrating surveys to understanding star formation, planet engulfments, and beyond.

\section{Acknowledgements}

TN \& KH have been partially supported by a TDA/Scialog  (2018-2020) grant funded by the Research Corporation and a TDA/Scialog grant (2019-2021) funded by the Heising-Simons Foundation. TN \& KH acknowledge support from the National Science Foundation grant AST-1907417. YST is grateful to be supported by the NASA Hubble Fellowship grant HST-HF2-51425.001 awarded by the Space Telescope Science Institute. KH is partially supported through the Wootton Center for Astrophysical Plasma Properties funded under the United States Department of Energy collaborative agreement DE-NA0003843. APJ acknowledges support from a Carnegie Fellowship and the Thacher Research Award in Astronomy. HK  acknowledges  support  from  the  DOE  CSGF  under  grant  number DE-FG02-97ER25308. The authors thank Carnegie Observatory for granting us the observing time to conduct this study.

This research has made use of the SIMBAD database, operated at CDS, Strasbourg, France \citep{simbad2000} and NASA’s Astrophysics Data System Bibliographic Services. This work has made use of data from the European Space Agency mission \textit{Gaia} (\url{https://www.cosmos.esa.int/gaia}), processed by the Gaia Data Processing and Analysis Consortium (DPAC, \url{https://www.cosmos.esa.int/web/gaia/dpac/consortium}). Funding for the DPAC has been provided by national institutions, in particular the institutions participating in the Gaia Multilateral Agreement.

\facilities{Magellan/Clay Telescope, Simbad}

\software{{\tt astropy} \citep{astropy:2013, astropy:2018},
{\tt NumPy} \citep{2020NumPy-Array}, {\tt iPython} \citep{ipython}, {\tt Matplotlib} \citep{matplotlib}, {\tt Galpy} \citep{Galpy}, {\tt SciPy} \citep{2020SciPy-NMeth}, {\tt Photutils} \citep{phot_utils}, {\tt BACCHUS} \citep{bacchus2016}, {\tt CarPy} \citep{Kelson03}, {\tt topcat} \citep{topcat}, {\tt iSpec} \citep{ispec}}

\startlongtable
\begin{deluxetable}{ll}

\tablecaption{Atomic Data References\label{tab:line_refs}}

\tablehead{Reference Key & Reference}
\startdata
1968PhFl...11.1002W & {\cite{1968PhFl...11.1002W}} \\
1969AA.....2..274G & {\cite{1969AA.....2..274G}} \\
1970AA.....9...37R & {\cite{1970AA.....9...37R}} \\
1970ApJ...162.1037W & {\cite{1970ApJ...162.1037W}} \\
1980AA....84..361B & {\cite{1980AA....84..361B}} \\
1980ZPhyA.298..249K & {\cite{1980ZPhyA.298..249K}} \\
1982ApJ...260..395C & {\cite{1982ApJ...260..395C}} \\
1982MNRAS.199...21B & {\cite{1982MNRAS.199...21B}} \\
1983MNRAS.204..883B & {\cite{1983MNRAS.204..883B}} \\
1984MNRAS.207..533B & {\cite{1984MNRAS.207..533B}} \\
1984MNRAS.208..147B & {\cite{1984MNRAS.208..147B}} \\
1984PhST....8...84K & {\cite{1984PhST....8...84K}} \\
1985AA...153..109W & {\cite{1985AA...153..109W}} \\
1985JQSRT..33..307D & {\cite{1985JQSRT..33..307D}} \\
1986JQSRT..35..281D & {\cite{1986JQSRT..35..281D}} \\
1986MNRAS.220..289B & {\cite{1986MNRAS.220..289B}} \\
1989AA...208..157G & {\cite{1989AA...208..157G}} \\
1989ZPhyD..11..287C & {\cite{1989ZPhyD..11..287C}} \\
1990JQSRT..43..207C & {\cite{1990JQSRT..43..207C}} \\
1991JPhB...24.3943H & {\cite{1991JPhB...24.3943H}} \\
1992AA...255..457D & {\cite{1992AA...255..457D}} \\
1993AAS...99..179H & {\cite{1993AAS...99..179H}} \\
1993JPhB...26.4409B & {\cite{1993JPhB...26.4409B}} \\
1993PhyS...48..297N & {\cite{1993PhyS...48..297N}} \\
1995JPhB...28.3485M & {\cite{1995JPhB...28.3485M}} \\
1996PhRvL..76.2862V & {\cite{1996PhRvL..76.2862V}} \\
1998PhRvA..57.1652Y & {\cite{1998PhRvA..57.1652Y}} \\
1999ApJS..122..557N & {\cite{1999ApJS..122..557N}} \\
2000MNRAS.312..813S & {\cite{2000MNRAS.312..813S}} \\
2003ApJ...584L.107J & {\cite{2003ApJ...584L.107J}} \\
2006JPCRD..35.1669F & {\cite{2006JPCRD..35.1669F}} \\
2006JPhB...39.2861Z & {\cite{2006JPhB...39.2861Z}} \\
2007AA...472L..43B & {\cite{2007AA...472L..43B}} \\
2007PhyS...76..577L & {\cite{2007PhyS...76..577L}} \\
2009A
2009AA...497..611M & {\cite{2009AA...497..611M}} \\
2009JPhB...42r5002K & {\cite{2009JPhB...42r5002K}} \\
2013ApJS..205...11L & {\cite{2013ApJS..205...11L}} \\
2013ApJS..208...27W & {\cite{2013ApJS..208...27W}} \\
2014ApJS..211...20W & {\cite{2014ApJS..211...20W}} \\
2014MNRAS.441.3127R & {\cite{2014MNRAS.441.3127R}} \\
ABH & {\cite{ABH}} \\
AMS & {\cite{AMS}} \\
AMb & {\cite{AMb}} \\
APH & {\cite{APH}} \\
APR & {\cite{APR}} \\
AS & {\cite{AS}} \\
ASa & {\cite{ASa}} \\
ASb & {\cite{ASb}} \\
ATJL & {\cite{ATJL}} \\
Astrophysical & {\cite{K07, K09}}\footnote{There are four lines with this designation. Two Fe lines and two V lines. The Fe lines come from \cite{K07} and the V lines from \cite{K09}. The log$gf$ values were astrophysically calibrated.}\\
BBC & {\cite{BBC}} \\
BBEHL & {\cite{BBEHL}} \\
BDMQ & {\cite{BDMQ}} \\
BGF & {\cite{BGF}} \\
BGHL & {\cite{BGHL}} \\
BGHR & {\cite{BGHR}} \\
BGKZ & {\cite{BGKZ}} \\
BHN & {\cite{BHN}} \\
BIEMa & {\cite{BIEMa}} \\
BIEMb & {\cite{BIEMb}} \\
BIPS & {\cite{BIPS}} \\
BK & {\cite{BK}} \\
BKK & {\cite{BKK}} \\
BKM & {\cite{BKM}} \\
BKP,BKM & {\cite{BKP,BKM}} \\
BKPb & {\cite{BKPb}} \\
BKor & {\cite{BKor}} \\
BL & {\cite{BL}} \\
BLNP & {\cite{BLNP}} \\
BLQS & {\cite{BLQS}} \\
BQR & {\cite{BQR}} \\
BQZ & {\cite{BQZ}} \\
BRD & {\cite{BRD}} \\
BSB & {\cite{BSB}} \\
BSScor & {\cite{BSScor}} \\
BWL & {\cite{BWL}} \\
BWL,BK & {\cite{BWL,BK}} \\
BXPNL & {\cite{BXPNL}} \\
Bar,BBC & {\cite{Bar,BBC}} \\
CB & {\cite{CB}} \\
CBcor & {\cite{CBcor}} \\
CC & {\cite{CC}} \\
CCout & {\cite{CCout}} \\
CM & {\cite{CM}} \\
CRC & {\cite{CRC}} \\
CSE & {\cite{CSE}} \\
DCWL & {\cite{DCWL}} \\
DHL & {\cite{DHL}} \\
DHWL & {\cite{DHWL}} \\
DIKH & {\cite{DIKH}} \\
DLSC & {\cite{DLSC}} \\
DLSSC & {\cite{DLSSC}} \\
DLW & {\cite{DLW}} \\
DLa & {\cite{DLa}} \\
DLb & {\cite{DLb}} \\
DSJ & {\cite{DSJ}} \\
DSLa & {\cite{DSLa}} \\
DSLb & {\cite{DSLb}} \\
DSLc & {\cite{DSLc}} \\
ESTM & {\cite{ESTM}} \\
FDLP & {\cite{FDLP}} \\
FMW & {\cite{FMW}} \\
GARZ & {\cite{GARZ}} \\
GC & {\cite{GC}} \\
GESB79b & {\cite{GESB79b}} \\
GESB82c & {\cite{GESB82c}} \\
GESB82d & {\cite{GESB82d}} \\
GESB86 & {\cite{GESB86}} \\
GESG12 & {\cite{GESG12}} \\
GESHRL14 & {\cite{GESHRL14}} \\
GESMCHF & {\cite{GESMCHF}} \\
GESOP & {\cite{GESOP}} \\
GHLa & {\cite{GHLa}} \\
GHR & {\cite{GHR}} \\
GHcor & {\cite{GHcor}} \\
GKOPK & {\cite{GKOPK}} \\
GKOa & {\cite{GKOa}} \\
GKOb & {\cite{GKOb}} \\
GNEL & {\cite{GNEL}} \\
GUES & {\cite{GUES}} \\
HLB & {\cite{HLB}} \\
HLGBW & {\cite{HLGBW}} \\
HLGN & {\cite{HLGN}} \\
HLL & {\cite{HLL}} \\
HLSC & {\cite{HLSC}} \\
IAN & {\cite{IAN}} \\
ILW & {\cite{ILW}} \\
JMG & {\cite{JMG}} \\
K03 & {\cite{K03}} \\
K04 & {\cite{K04}} \\
K06 & {\cite{K06}} \\
K07 & {\cite{K07}} \\
K08 & {\cite{K08}} \\
K09 & {\cite{K09}} \\
K10 & {\cite{K10}} \\
K11 & {\cite{K11}} \\
K12 & {\cite{K12}} \\
K13 & {\cite{K13}} \\
K14 & {\cite{K14}} \\
K75 & {\cite{K75}} \\
K99 & {\cite{K99}} \\
KG & {\cite{KG}} \\
KK & {\cite{KK}} \\
KP & {\cite{KP}} \\
KR & {\cite{KR}} \\
KSG & {\cite{KSG}} \\
KZB & {\cite{KZB}} \\
KZBa & {\cite{KZBa}} \\
LAW & {\cite{LAW}} \\
LBS & {\cite{LBS}} \\
LCG & {\cite{LCG}} \\
LCV & {\cite{LCV}} \\
LD & {\cite{LD}} \\
LD-HS & {\cite{LD-HS}} \\
LDLS & {\cite{LDLS}} \\
LGWSC & {\cite{LGWSC}} \\
LGb & {\cite{LGb}} \\
LMW & {\cite{LMW}} \\
LN & {\cite{LN}} \\
LNAJ & {\cite{LNAJ}} \\
LNWLX & {\cite{LNWLX}} \\
LSC & {\cite{LSC}} \\
LSCI & {\cite{LSCI}} \\
LSCW & {\cite{LSCW}} \\
LV & {\cite{LV}} \\
LWCS & {\cite{LWCS}} \\
LWG & {\cite{LWG}} \\
LWHS & {\cite{LWHS}} \\
LWST & {\cite{LWST}} \\
LWa & {\cite{LWa}} \\
MC & {\cite{MC}} \\
MFW & {\cite{MFW}} \\
MIGa & {\cite{MIGa}} \\
MIGb & {\cite{MIGb}} \\
MRB & {\cite{MRB}} \\
MRW & {\cite{MRW}} \\
MULT & {\cite{MULT}} \\
MW & {\cite{MW}} \\
MWRB & {\cite{MWRB}} \\
NG & {\cite{NG}} \\
NHEL & {\cite{NHEL}} \\
NI & {\cite{NI}} \\
NIJL & {\cite{NIJL}} \\
NIST10 & {\cite{NIST10}} \\
NWL & {\cite{NWL}} \\
NZL & {\cite{NZL}} \\
OK & {\cite{OK}} \\
PGBH & {\cite{PGBH}} \\
PGHcor & {\cite{PGHcor}} \\
PGK & {\cite{PGK}} \\
PK & {\cite{PK}} \\
PN & {\cite{PN}} \\
PQB & {\cite{PQB}} \\
PQWB & {\cite{PQWB}} \\
PRT & {\cite{PRT}} \\
PST & {\cite{PST}} \\
PSa & {\cite{PSa}} \\
PTP & {\cite{PTP}} \\
PV & {\cite{PV}} \\
QPB & {\cite{QPB}} \\
QPBM & {\cite{QPBM}} \\
RHL & {\cite{RHL}} \\
RPU & {\cite{RPU}} \\
RSa & {\cite{RSa}} \\
RU & {\cite{RU}} \\
RW & {\cite{RW}} \\
S & {\cite{S}} \\
S-G,BBC & {\cite{S-G,BBC}} \\
SCRJ & {\cite{SCRJ}} \\
SDL & {\cite{SDL}} \\
SDcor & {\cite{SDcor}} \\
SEN & {\cite{SEN}} \\
SG & {\cite{SG}} \\
SK & {\cite{SK}} \\
SLS & {\cite{SLS}} \\
SLb & {\cite{SLb}} \\
SLd & {\cite{SLd}} \\
SN & {\cite{SN}} \\
SPN & {\cite{SPN}} \\
SR & {\cite{SR}} \\
Si2-av1 & {\cite{Si2-av1}} \\
Sm & {\cite{Sm}} \\
T & {\cite{T}} \\
T83av & {\cite{T83av}} \\
TB & {\cite{TB}} \\
VGH & {\cite{VGH}} \\
WBW & {\cite{WBW}} \\
WBb & {\cite{WBb}} \\
WGTG & {\cite{WGTG}} \\
WL & {\cite{WL}} \\
WLN & {\cite{WLN}} \\
WLSC & {\cite{WLSC}} \\
WLa & {\cite{WLa}} \\
WM & {\cite{WM}} \\
WSG & {\cite{WSG}} \\
WSL & {\cite{WSL}} \\
WSM & {\cite{WSM}} \\
WV & {\cite{WV}} \\
Wa & {\cite{Wa}} \\
Wc & {\cite{Wc}} \\
XJZD & {\cite{XJZD}} \\
XSCL & {\cite{XSCL}} \\
XSQG & {\cite{XSQG}} \\
ZLLZ & {\cite{ZLLZ}} \\
ZZZ & {\cite{ZZZ}} \\
\enddata

\end{deluxetable}
\newpage

\appendix
\section{Estimating 3D separations for wide binaries}
\label{sec:3d_WB}
Consider a pair of stars with true 3D separation $r_{\rm 3D}$. If the pair is viewed along a random viewing angle, the probability distribution of its projected separation $r_{\rm 2D}$ can be calculated via a straightforward geometric argument \citep[e.g.][]{r2dr3d_projection}:
\begin{equation}
    p\left(r_{\rm 2D}|r_{\rm 3D}\right)=\frac{1}{r_{\rm 3D}}\frac{r_{\rm 2D}/r_{\rm 3D}}{\sqrt{1-\left(r_{\rm 2D}/r_{\rm 3D}\right)^{2}}}.
    \label{eq:r2dr3d}
\end{equation}
For the pairs in our sample suspected to be binaries, we have a precise measurement of $r_{\rm 2D}$ and wish to constrain $r_{\rm 3D}$. By Bayes' rule, the probability distribution of $r_{\rm 3D}$ given a measurement or $r_{\rm 2D}$ depends both on the measured $r_{\rm 2D}$, and on the prior on $r_{\rm 3D}$.
\begin{equation}
    p\left(r_{\rm 3D}|r_{\rm 2D}\right)\propto p\left(r_{\rm 2D}|r_{\rm 3D}\right)p\left(r_{\rm 3D}\right).
    \label{eq:r3dr2d}
\end{equation}

\begin{figure}
    \centering
    \includegraphics[width=0.6\columnwidth]{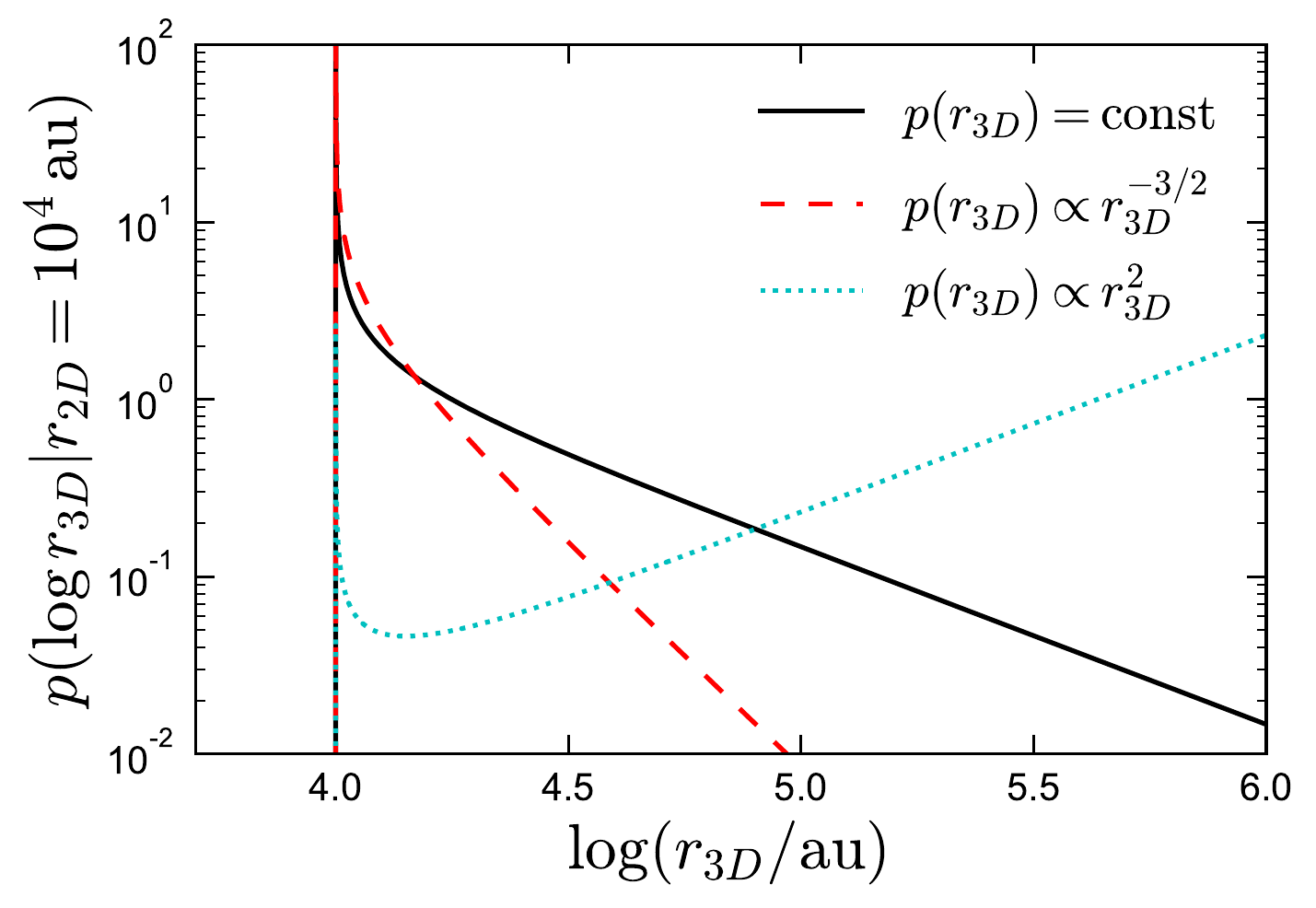}
    \caption{Conditional probability distribution of the logarithmic 3D separation $r_{\rm 3D}$, given a measured projected separation of $r_{\rm 2D}=10^4$\,AU. We compare distribution for three different choices of the prior on the intrinsic 3D separation distribution, $p(r_{\rm 3D})$. Dashed red line shows the prior we adopt for pairs suspected to be wide binaries. For both this choice and a flat prior (black line), $p\left(\log r_{\rm 3D}|r_{\rm 2D}\right)$ is sharply peaked at $r_{\rm 2D}$; that is, the most likely 3D separation is within a factor of a few of the projected separation. Dotted cyan line shows  $p(r_{\rm 3D})\propto r_{\rm 3D}^2$, as is expected for random chance alignments. In this case, the probability mass is dominated by pairs with $r_{\rm 3D} \gg r_{\rm 2D}$; that is, widely separated pairs viewed almost end-on.}
    \label{fig:p_r3_given_r2d}
\end{figure}

The separation distribution of wide binaries is well-constrained observationally to be $p\left(r_{\rm 3D}\right)\sim r_{\rm 3D}^{-3/2}$ at $r_{\rm 3D} > 500$\,AU \citep[e.g.][]{Andrews_2017, Elbadry_and_Rix_2018}, so we adopt this as the prior for the 3D separation distribution of pairs thought to be binaries. In this case, Equation~\ref{eq:r3dr2d} becomes
\begin{equation}
    p\left(r_{\rm 3D}|r_{\rm 2D}\right)=\begin{cases}
\frac{2}{\sqrt{\pi}}\frac{\Gamma\left(7/4\right)}{\Gamma\left(5/4\right)}\frac{1}{r_{\rm 3D}}\frac{\left(r_{\rm 2D}/r_{\rm 3D}\right)^{5/2}}{\sqrt{1-\left(r_{\rm 2D}/r_{\rm 3D}\right)^{2}}}, & r_{\rm 3D}>r_{\rm 2D}\\
0 & {\rm otherwise}
\end{cases},
    \label{eq:r3dr2d_with_prior}
\end{equation}
where $\Gamma$ represents the Gamma function, and the coefficient is derived from the normalization condition. The red line in Figure~\ref{fig:p_r3_given_r2d} shows this distribution, with $r_{\rm 2D}=10^4$\,AU adopted for concreteness.\footnote{For better visual interpretability, we plot $p\left(\log r_{{\rm 3D}}|r_{{\rm 2D}}\right)=\ln10\times r_{3D}\times p\left(r_{3{\rm D}}|r_{{\rm 2D}}\right)$.} For this distribution, the median value of $r_{\rm 3D}/r_{\rm 2D}$ is 1.12, and the middle 68.2\% range is (1.01, 1.61). Given a measurement of $r_{\rm 2D}$, we thus report $r_{\rm 3D}=1.12_{-0.11}^{+0.49}\times r_{\rm 2D}$ if the pair is suspected to be a binary.

Of course, other choices are possible for the prior $p(r_{\rm 3D})$ when the separation exceeds what we expect from the binary population. The black line in Figure~\ref{fig:p_r3_given_r2d} corresponds to a ``flat'' prior, $p(r_{\rm 3D})=\rm const$. Such a separation distribution might be expected for conatal pairs that are no longer bound, if the birth and dissolution rates of such pairs are constant. In this case, the median and 1$\sigma$ range in $r_{\rm 3D}$ would be $r_{\rm 3D}=1.41_{-0.38}^{+2.63}\times r_{\rm 2D}$, still implying that most pairs have 3D separations within a factor of a few of their 2D projected separation.

Finally, the dotted cyan line in Figure~\ref{fig:p_r3_given_r2d} shows the results of assuming $p(r_{\rm 3D})\propto r_{\rm 3D}^2$. This is the distribution expected for random pairings of isotropically distributed stars. In this case, the probability distribution is normalizable only if we assume that there are no pairs beyond a 3D separation $r_{\rm max}$. The conditional probability distribution in this case is 
\begin{equation}
    p\left(r_{\rm 3D}|r_{\rm 2D}\right)=\begin{cases}
\frac{1}{\sqrt{\left(1-r_{\rm 2D}^{2}/r_{\rm 3D}^{2}\right)\left(r_{{\rm max}}^{2}-r_{\rm 2D}^{2}\right)}}, & r_{\rm 3D}>r_{\rm 2D}\\
0 & {\rm otherwise}
\end{cases},
    \label{eq:r3dr2d_with_r2prior}
\end{equation}
which approaches a constant value at $r_{\rm 3D} \gg r_{\mathrm{2D}}$. Intuitively, what happens in this case is that the larger number of possible pairs at large $r_{\rm 3D}$ (which scales as $r_{\rm 3D}^2$) exactly compensates for the smaller number of sightlines along which a pair can be viewed to have a given projected separation (which scales as $r_{\rm 3D}^{-2}$). In Figure~\ref{fig:p_r3_given_r2d}, we adopt $r_{\rm max}= 10^6$\,AU.


To summarize, we calculate constraints on the 3D separation of pairs thought to be wide binaries based on their projected 2D separation using Equation~\ref{eq:r3dr2d_with_prior}. This yields a median and 1$\sigma$ range of $r_{\rm 3D}=1.12_{-0.11}^{+0.49}\times r_{\mathrm{2D}}$, which at small $r_{\mathrm{2D}}$ is much more constraining than the constraint on $r_{\rm 3D}$ that considers the distances to both stars independently (``method 1''). The primary assumption of this constraint is that the separation distribution of wide binaries falls off as $p(r_{\mathrm{3D}})\propto r_{\mathrm{3D}}^{-3/2}$. It is not appropriate for pairs that are not binaries (unbound comoving pairs or chance alignments), because their separation distribution is expect to {\it increase} with separation due to the different priors. We thus only apply this correction to pairs with separation $< 2 \times 10^5\,$AU.

\section{Line List references}
A large part of this work is made possible by numerous heroic efforts by various groups who perform the pain-stacking tasks of curating/calibrating the atomic line list used in BACCHUS. Those efforts are unfortunately under appreciated in the literature. As such, albeit long, we decide to include the references to all original sources of the line list adopted in this study in Table \ref{tab:line_refs}.

\bibliography{wide_binary}

\end{document}